\documentstyle[aps]{revtex}
\narrowtext
\begin{document}

\title{Quantum Kinetic Theory I: A Quantum Kinetic Master Equation for
 Condensation of a weakly interacting Bose gas without a trapping potential}
\author{C.W.~Gardiner$^{1}$ and P.~Zoller$^2$}
\address{$^1$ Physics Department, Victoria University, Wellington, New Zealand}
\address{$^2$ Institut f{\"u}r Theoretische Physik,
Universit{\"a}t Innsbruck, 6020 Innsbruck, Austria}

\maketitle

\begin{abstract}
A Quantum Kinetic Master Equation (QKME) for bosonic atoms is formulated. It is
a quantum stochastic equation for the
kinetics of a dilute quantum Bose gas, and describes the behavior
and formation of Bose condensation.
The key assumption in deriving the QKME is a Markov approximation for the
atomic collision terms.
In the present paper the basic structure of the theory is developed,  and
approximations are stated and justified to delineate the region of validity of
the theory.
Limiting cases of the QKME include the Quantum Boltzmann master equation and
the Uehling-Uhlenbeck equation,
as well as an equation analogous to the Gross-Pitaevskii
equation.
\end{abstract}


\narrowtext
\section{Introduction}
Recent observations of Bose-Einstein condensation (BEC) in a dilute
gas of magnetically trapped alkali atoms %
\cite{JILA,Rice,MIT,newmit,JILAexcitation,MITexcitation},
and in
excitonic systems
\cite{excitons}
has stimulated new theoretical
efforts to describe the dynamics and signatures of weakly interacting
Bose gases. In contrast to superfluid Helium, which for many years has
been the only experimental example of BEC, these weakly interacting
Bose gases are much more amenable to a theoretical analysis.  Recent
theoretical work has focused on a description of BEC in a trapping
potential and dynamics of formation of a Bose condensate
\cite{KaganKinetics,KagSvis,KaganShlyapnikovWalraven,evaporativecooling,%
KaganExpansion,Stoof91,Stoof94,Stoof96,Levich,Snoke,Semikoz,Goldman,Siggia,%
Lovelace,Burnett95a,Burnett95b,Ballagh,Edwards1996a,Edwards1996b,Baym,%
Stringari96a,Stringari96b,Stringari96c,YouHolland,HollandCooper,%
HollandBoltzmann,%
LewensteinYou,Youexcitationspectrum,DumCastin,Greene,%
excitationspectrum,Griffin96}.

In this paper we want to develop techniques to describe the behavior
and formation of the Bose condensate in which we will combine the
simplicity of a quantum stochastic methods used in quantum optics
\cite{Gardiner}
with a realistic treatment of the interatomic interactions which are known
to play a major role in the dynamics of the condensing system
\cite{Trento}.
Such a description must necessarily involve aspects of a kinetic theory as
well as quantum mechanics.

The first realistic formulation of kinetic
theory was the Boltzmann Equation (BE) whose usefulness and success to this
day is universally accepted. The so-called Quantum Boltzmann or
Uehling-Uhlenbeck equation (QBE)
\cite{U-U} introduces corrections for quantum statistics into the
Boltzmann collision term, but since it deals only with one-particle
distribution functions we cannot expect this equation to give a
realistic treatment of the quantum mechanical aspects which must occur
in a Bose condensate.

A different description is provided by the time
dependent Gross-Pitaevskii equation (GP equation)
\cite{GPoldies}
which can be viewed as an equation for the
condensate wave function (order parameter for the Bose condensate).
This equation is clearly a simplified description in that it includes
no quantum fluctuations, or thermal or irreversible effects, but it
may well be valid in the situation of a large number of condensate
particles. Both of these equations contain essential aspects of the
problem we are studying. However, in practice the process of creating
a Bose condensate in a trap by means of evaporative cooling starts in a
regime covered by a kinetic equation and finishes in a regime where the
GP is thought to be valid. What is needed is unified description which
covers the whole range --- a true quantum kinetic theory. In this
paper we will develop a {\em quantum kinetic master equation} (QKME)
which is not a mere modification of the BE but extends its philosophy
into quantum mechanics and derives a {\em quantum stochastic equation}
of the kind similar to those derived in quantum optics.

The QKME which we derive is thus a consistent extension of the
philosophy of the BE to a regime in which quantum mechanical effects are
not a minor correction to classical results, but play a major role in
the full description of the system.

In applying the QKME to the problem of Bose condensation, although
quantum mechanics plays major role, the significant quantum aspects
are restricted to a few modes, the remaining modes being able to be
describe in a manner similar to the classical Boltzmann equation.  This is a
situation
which is very familiar in quantum optics, where we very often find
that it is necessary to describe a few optical modes fully quantum
mechanically, while treating the rest of the systems as a heat bath.
However, the methods necessary for quantum kinetic theory contain
considerably more technical difficulty, largely because of the
interaction between the atoms. The nearest optical analogy is
multimode propagation in a medium with a strong nonlinear refractive
index \cite{DrummondCarter}.
\section{The need for a quantum kinetic theory}\label{Sect. 6.1}
Our goal is to derive a quantum kinetic master equation for a gas of
$N$ weakly interacting Bose atoms. The master equation is formulated
as an evolution equation for the $N$-atom density matrix, giving a
quantum mechanical generalization of the Quantum Boltzmann equation.
This provides a fully quantum mechanical description of the kinetics
of a Bose gas, including the regime of Bose condensation. In
particular, such an equation is capable of describing the formation of
the Bose condensate.  The quantum mechanical processes which must be
included in this treatment are the atomic motion (transport) and
coherent and incoherent interactions (collisions) between the atoms.
In this section we give a qualitative overview over the main results
to be derived in later sections.

\subsection{Essential elements of the theory}
There are three principal elements, as follows.
\subsubsection{Coherent and incoherent dynamics}
The derivation of a quantum master equation implies that the system dynamics
can
divided into into {\em coherent} and {\em incoherent} physical
processes, where the incoherent dynamics is modeled as a quantum
stochastic Markovian process. This builds on ideas, which have been
developed and applied successfully in quantum optics as a method of
describing coherent and incoherent processes simultaneously (e.g. in
laser theory).

In a weakly interacting Bose gas, we will treat the collisions between
the atoms resulting in a {\it large momentum transfer} as random and weak
incoherent processes, responsible for the ``noise'' in the system.  On
the other hand, forward or near forward scattering is, if the
wavelength is sufficiently long, largely a coherent process, giving
rise to dispersive effects.

\subsubsection{Dynamics of the condensate}
The (coherent) dynamics of the Bose condensate is treated explicitly
and separate from the description of the non-condensed modes which
play the role of noise and feeding terms for the Bose condensate.
Again, this is guided by laser theory where the coherent laser mode is
separated from the weakly populated incoherent modes which are
typically treated as a heat bath.

\subsubsection{The ``cell'' description}
The theory is formulated in terms of a quantum mechanical {\em phase space
description} for with coarse grained position and momentum variables, based
on division of phase space into ``cells'' of volume $h^3$. This seems
to be the most natural formulation, since collisions are best
described in terms of momentum, whereas transport is best understood
in terms of position.  We will introduce this formulation by means of a wavelet
description, which gives an {\em exact} exact description of the dynamics;
approximations appear as a result of the procedure for describing the kinetics,
and the conditions for the validity of our approximations then put some
conditions on the precise nature of the cells chosen.

\subsection{The scope of this paper}
In developing the physical picture, there are three main stages through which
we must proceed in order to achieve a full quantum kinetic theory applicable to
the present experiments on Bose-Einstein condensation.
\begin{trivlist}
\item[i)]
It is necessary first to develop the theory for the weakly interacting Bose gas
with no trapping potential, whose stationary solution is well approximated by
an ideal Bose gas.
\item[ii)]
The next stage is to take account of the modifications (these may be very
significant) to the dynamics, which are induced by the presence of a large
proportion of atoms being in the condensate.  This can be developed in terms of
Bogoliubov quasiparticles
\cite{quasiparticles,Lifshitz},
or more generally.
\item[iii)]
Finally, we introduce the trapping potential.  There are two extreme situations
to consider.  If the trap is very broad, then the situation is qualitatively
not very different from the case of no trap, though the appropriate
modifications can be quite intricate.  In the case of a very tight trap, we
have a completely different treatment, based on the kinetics of transitions
between different trap levels.  There will also be an intermediate regime, in
which it is relevant to treat only the lower trap levels explicitly, while
upper levels are thermalized.
\end{trivlist}
This paper deals with i) only; ii) and iii) will be treated in our second paper
QKII.  Thus this paper develops all the basic ideas and methodology, while QKII
will develop the refinements necessary to deal with strong condensation with
and without a trapping potential.
\subsubsection{The physical picture}
This phase space description is one in which
particles are represented as wave packets interacting with each other.
From the point of view of wavepackets there are the following processes:
{\em Transport:} that is the motion through space of a wavepacket.
However we also understand that from a wavepacket point of view,
transport is a process by which the wavepacket moves with unchanged
shape---the wave\-pack\-et spreading inherent in the quantum mechanics
of particle motion is not itself viewed as a part of transport.
{\em Wavepacket spreading} occurs in addition to what one would
normally call transport, and must contribute to the production of
coherence which can arise in a Bose condensed system.  It is purely
quantum mechanical, but arises of course from exactly the same quantum
mechanical source as transport, as defined here. Motion in a trap is readily
included in these arguments to the extent the trapping potential can be
considered as slowly varying.
{\em Collisions:} here we mean localized events in which the
momentum of wave\-pack\-ets will change instantaneously, with a
consequent change in the wavefunction.  The localization cannot of
course be exact, because wavepackets themselves are extended objects.


\subsubsection{The size of the cells}
Quantum mechanics itself does not give any preferred size of the
momentum space or coordinate space cells separately.  Physically, we
can characterize the choice of cells by a wavenumber $ \Delta$
(or a cell size $ l_c=\pi /\Delta$
(compare Fig.~1), and
this will naturally provide a ``cut'' in {\em momentum space}; changes of
momentum greater than $ \hbar\Delta$ are seen as collisions, while
smaller changes are seen as changes which take place in coordinate
space.  The value of $ \Delta$ is constrained by two considerations.
\begin{enumerate}
\begin{enumerate}
\item We want to have a description in terms of ``energy levels'' within a box
inside each cell.  This means that the wavefunction of a particular particle
must maintain its coherence over the length $ l_c$ of the cell.  The coherence
length which arises from collisions should normally be of the same order of
magnitude as the mean free path $\lambda_{\rm mfp}= v\tau$, where $ v$ is the 
speed of
the particle,and $ \tau$ is the mean time between collisions.  Thus this
requirement yields the condition $ l_c \ll \lambda_{\rm mfp}$.

\item We will be wanting to develop a Markovian description of collisions,
which depends on the determination of a characteristic decoherence time,
which is essentially the bandwidth of frequencies of particles in the system,
that is, {\em thermal correlation time} $ \hbar/2\pi k T $.    This should be
much less than the the characteristic time for evolution of the the phase-space
density, and this yields the condition that the mean free path
$ \lambda_{\rm mfp} \gg \lambda_{T}= h/\sqrt{2mkT}$, which is the wavelength of
a typical particle, known as the {\em thermal wavelength}.

\item As well as the condition that the bandwidth of frequencies must be 
sufficiently 
large, we must also have a sufficient density of states to ensure that the 
correlation function inherent does indeed become very small for times greater 
than the correlation time.
Essentially this requires that the maximum energy
difference between cells should be much less than $ kT$, leading to the
requirement $ l_c \gg l_T\equiv h/\sqrt{2mkT}$, that is the cells much be very
much larger than the thermal wavelength. 
\item The method is based on a requirement that correlations which involve 
momentum differences greater than $ {\bf \Delta}$ are negligible. This means 
that the ``collisions'', that is, processes which cause a momentum change 
larger than $ |{\bf \Delta}|$, are seen as inducing incoherence.  The 
derivation of the QKME which we use does this by means of a projection 
formalism, and the condition for the validity of the procedure is found to be 
\begin{eqnarray}\label{AA}
a\lambda_{\rm mfp}\lambda_T \ll l_c^3
\end{eqnarray}
where $ a$ is the scattering length for the interaction between the particles.
\item There is also a {\em weak condensation condition} for the validity of the
methods
used in this paper, and this can be written in terms of the cell size $ l_c$,
the scattering length $ a$, and the particle density $ \rho $  as
$ l_c \gg  \sqrt{\pi/8a\rho}$.
\end{enumerate}
\end{enumerate}

These criteria are easily met, for example in the experiment of
\cite{MIT}
one finds that
$ \lambda= 0.4619{\rm m}$, while the thermal wavelength is
$ l_T = 4.8\times10^{-7}{\rm m}$.

On the other hand in that experiment the trap itself has a size of the order of
magnitude as  $ l_T$.  Thus to describe the behavior of the wavefunction
within the trap,
one would want $ l_c$ to be finer than this.  This first paper does not deal
with trapping potentials, but we will show how to overcome this difficulty in
the QKII.  The essence of the solution is that it is not necessary to
have the same size cells for all ranges of momentum---they can be finer in
space at the lower momentum ranges.

Figs.~2  and 3 illustrate the phase space description and the various
physical processes. The figures show the phase space cells
corresponding to different momentum bands ${\bf K}$ with width ${\bf
\Delta}$ and position cells
${\bf r}$. A state of the system is given by specifying a set of
occupation numbers $\{ n({\bf K},{\bf r}) \}$ of the cells (Fig.~2).  Transport
and wave packet spreading corresponds to coherent quantum processes
connecting cells in a given momentum band (horizontal arrows in Fig.~3a). Note
that these coherent processes do not change a given momentum
distribution
${\bf N}({\bf K})=(N({\bf K}_1),N({\bf K}_2),\ldots)$
with $N({\bf K}) = \sum_{\bf r} n({\bf K},{\bf r})$ the number of
atoms in a momentum band. In a similar way, coherent processes
corresponding to a forward scattering, or motion in a slowly varying
trapping potential give rise to smooth shifts between the cells of
neighboring ${\bf K}$ (Fig.~3a and b).  Incoherent collision processes between 
two
atoms, on the other hand, are associated with a ``quantum jump like''
momentum change ${\bf K}_1, {\bf K}_2 \rightarrow {\bf K}_3,{\bf K}_4$
between widely separated in momentum bands (however, essentially
confined to within a single spatial cell $\bf r$) (see Fig.~3c).  These 
transitions
conserve momentum,
\begin{equation}
{\bf K}_1 + {\bf K}_2 = {\bf K}_3 + {\bf K}_4
+{\cal O}({\bf
\Delta}),
\end{equation}
up to an uncertainty $\Delta$ associated with the width of the band, and energy
\begin{equation}
E_1 + E_2 = E_3 + E_4 \quad \mbox{($E_i = \hbar^2 {\bf K}^2/2 m$)}.
\end{equation}
Collisions cause transitions between different
momentum distributions
\begin{eqnarray}
{\bf N}({\bf K}) &=& (\ldots,n_1,n_2,n_3,n_4,\ldots)
\nonumber \\
&\rightarrow &
{\bf N}'({\bf K}) =(\ldots,n_1-1,n_2-1,n_3+1,n_4+1,\ldots)
\end{eqnarray}
which we write in the form
\begin{eqnarray}
{\bf N}({\bf K})\rightarrow {\bf N}'({\bf K})= {\bf N}({\bf K})+ {\bf e}.
\end{eqnarray}

\subsection{The quantum kinetic master equation}
The derivation of the master equation in Sec.~\ref{Sect. 6.3.3} below assumes
that the full quantum
coherence is kept for different spatial configurations (occupation
numbers of the phase space cells) within one $\bf K$-band, and the
assumption that atomic coherences between different momentum bands are
eliminated in the Born and Markov approximation.  The mathematical procedure of
deriving a master equation in Sec. {\ref{Sect. 6.2}} is based on
projecting the $N$-atom density operator $\rho$ on states associated
with a given momentum distributions ${\bf N}({\bf K})$. For a given cell size
there is a regime where this diagonally projected density matrix
$v_{\bf N}$ obeys a closed equation

\begin{mathletters}
\begin{eqnarray} \label{QKME}
 \dot v_{{\bf N}}( t) &=& -{ i\over\hbar}\left[ \sum\limits_{{\bf K}}
\int d^3{\bf x\,}{\left(\hbar {\bf K} \over 2m\right)}
\cdot {\bf j}_{{\bf K}}( {\bf x})
+
\sum\limits_{{\bf K}}\int d^3{\bf x\,} \psi _{{\bf K}}^{\dagger }({\bf
x})\left(-{\hbar ^2\nabla^2\over 2m}\right)
\psi_{{\bf K}}( {\bf x}) ,v_{{\bf N}}(t) \right]     \\
&& - {i\over2\hbar} \left[
\sum_{{\bf K}_1 \ne {\bf \,K}_2}
\left(U( {\bf K}_1,{\bf K}_2,{\bf K}_1,{\bf K}_2) +
U( {\bf K}_1,{\bf K}_2,{\bf K}_2,{\bf K}_1)\right) + 
\sum\limits_{{\bf K}_1}U( {\bf
K}_1,{\bf K }_1,{\bf K}_1,{\bf K}_1)
,v_{{\bf N}}(t)\right]
\\
 &&{ +{\pi \over\hbar
}{\sum\limits_{{\bf e}}}\delta (\Delta E({\bf e}))} \{ 2U({\bf
e})v_{{\bf N-e}}(t) U^{\dagger}({\bf e}) -
U^{\dagger}({\bf e})U({\bf e})v_{{\bf N}}(t)-v_{{\bf N}}(t)
U^{\dagger}({\bf e}) U({\bf e})\}.
\end{eqnarray}
\end{mathletters}
The first line in (\ref{QKME}) describes atomic motion between the
different phase space cells within a given $\bf K$--band. The first term is a
streaming term with current operator ${\bf j}_{{\bf K}}( {\bf x})$ (for a
formal  definition see (\ref{1.210001} below), and the second term
corresponds to wavepacket spreading (the field operator for a $\bf K$--band
will be defined in Eq.~(eq1.16)). The second line is (coherent) forward
scattering, and the last describes the redistribution by collisions according
to a $U({\bf e})$-operator, which is expressible in terms of the interaction
potential, and is explicitly defined in (\ref{eq1.24}). A detailed derivation
of the
equation will follow in
Sec.~IV below.


\subsubsection{Wavefunction stochastic differential equation (SDE)
interpretation}
It is interesting to interpret (\ref{QKME}) from the point of view of
evolution of a stochastic $N$-atom wavefunction where the wide angle collisions
are described as quantum jumps
\cite{trajectories}.
Note that (\ref{QKME}) can be written in the form
\begin{eqnarray} \label{masterequation}
 \dot v_{{\bf N}}( t) &=& -{ i\over\hbar} H_{\rm eff}  v _{{\bf N}}(t)
+
{ i\over\hbar}  v _{{\bf N}}(t) H_{\rm eff}  ^\dagger  \\
&&+
{2 \pi \over\hbar}
\sum_{{\bf e}}
\delta (\Delta E({\bf e})) \;
  U({\bf e})v_{{\bf N-e}}(t) U^{\dagger}({\bf e})  \; .
\end{eqnarray}
Let us define an atomic state vector $| \psi _{  {\bf N}  } (t)\rangle$ with
${\bf N} ({\bf K})$ a given configuration of occupation numbers.
In the time evolution of this state vector a collision is associated with an
instantaneous jump
\begin{equation}
| \psi _{  {\bf N}' + {\bf e}  } (t +dt)\rangle \propto  U({\bf e})
| \psi _{ {\bf N}   } (t )\rangle
\end{equation}
Here the $U$-matrix plays the role of the jump operator, connecting
configurations  ${\bf N} ({\bf K})  \rightarrow  {\bf N}({\bf K}+{\bf e})$. The
probability for a collision is in the time interval $(t,t+dt]$ is, given a
normalized statevector $| \psi _{  {\bf N} ({\bf K}) } (t)\rangle$ at time $t$,
by
\begin{eqnarray}
P_{(t,t+dt]} &=& {2 \pi \over \hbar} \sum_{{\bf e}} \delta (\Delta E({\bf e}))
|\!| U({\bf e})| \psi _{  {\bf N}  } (t)\rangle |\!|^2  \; dt\\
 &\equiv  & \sum_{{\bf e}} P_{(t,t+dt]} ({\bf e})
\end{eqnarray}
The time evolution between the collisions is governed by the nonhermitian
Hamiltonian $H_{\rm eff}$,
\begin{equation} \label{betweenjumps}
| \psi _{  {\bf N}({\bf K})} (t )\rangle =
e^{-i H_{\rm eff} t/\hbar}
| \psi _{ {\bf N} ({\bf K})  } (0)\rangle \; .
\end{equation}
Physically, (\ref{betweenjumps}) describes the streaming and forward
scattering between the jumps, and the nonhermitian is associated with loss due
to collisions. Note that $H_{\rm eff}$ preserves the distribution ${\bf N} ({
\bf K})$ (see Figs.~2a and b). A stochastic average over these quantum
trajectories give the density matrix, $v_{{\bf N}}(t)
= \langle \! \langle
| \psi _{  {\bf N}  } (t)\rangle  \langle \psi _{  {\bf N}  } (t) |
/ |\!|  \psi _{  {\bf N}  } (t)  |\!|^2   \rangle \! \rangle $.
\subsection{Limiting cases}
The Quantum Kinetic Master Equation (\ref{QKME}) includes as a limiting cases
both the Quantum Boltzmann master equation and the Uehling-Uhlenbeck equation,
as well as an equation analogous to the Gross-Pitaevskii
equation. These results will be derived in Sec.~\ref{Sect. 6.5.2}).
\subsubsection{Quantum Boltzmann master equation}
If we assume a {\em single spatial cell}, i.e.~the size of the system equals
the dimension of the system (compare Figs.~1 and 2), the transport terms
trivially disappear and Eq.~(\ref{QKME}) reduces
(Sec.~\ref{Sect. 6.5.2}) to  an equation which we shall call the {\em Quantum
Boltzmann master equation\/}:
\begin{eqnarray}\label{quantumBoltzmann}
\dot w_{\bf n} & = & -\frac{\pi}{\hbar^2}\sum_{1234}
\delta(\Delta E (1234)) |T_{1234}|^2
\nonumber\\
&&\qquad\qquad\quad\times\Big\{n_1n_2(n_3+1)(n_4+1)[w_{\bf n}-w_{\bf n+e}]
\nonumber \\
&& \qquad\qquad\quad+ (n_1+1)(n_2+1)n_3n_4[ w_{\bf n}-w_{{\bf n}-{\bf e}}]\Big
\}.
\;
\end{eqnarray}
This is a rate equation connecting the occupation probabilities of a given
configuration ${\bf n}\equiv {\bf N}$, $w_{\bf n}$, where  the factors $n_k+1$
reflect quantum statistics for the collisional transition rates.  It obviously
has a limited validity, but it is attractive because of its simplicity and ease
of simulation. \cite{Jaksch}.
\subsubsection{Uehling-Uhlenbeck equation}
For spatially inhomogeneous systems we derive from (\ref{QKME}) the
Uehling-Uhlenbeck equation (Sec.~(\ref{Sect. 6.4})). Let us define a single
particle distribution function $f_{\bf K} ({\bf x})$  where $\bf K$ labels a
momentum band and $\bf x$ is a (continuous) spatial coordinate (for a precise
definition see Eq.~(\ref{eq3.1})). We interpret $f_{\bf K} ({\bf x})$ as a
joint momentum position distribution function, similar to the Wigner function.
Assuming a factorization of the $N$--atom distributions
(an assumption not valid in the BEC regime) one obtains from (\ref{QKME}) the
kinetic equation
\begin{eqnarray}\label{UU}
{\partial \over \partial t} f_{{\bf K}} ({\bf x})) &\approx&
 \frac{\hbar {\bf K}\cdot \nabla_{{\bf x}}}m f_{\bf K}( {x})
 +\frac{2| u|^2}{h^2}\int\!\! \int\!\!\int d^3{\bf K}_2d^3{\bf K}_3d^3{\bf K}_4
\nonumber\\
&&\times \delta ( {\bf K}+{\bf K}_2-{\bf K}_3-{\bf K}_4)
\delta (\omega +\omega_2-\omega_3-\omega_4)
\nonumber \\
&&\times \{
f_{{\bf K}}( {\bf x}) f_{{\bf K}_2}( {\bf x} )
[ f_{{\bf K}_3}( {\bf x}) +1] [ f_{ {\bf K}_4}( {\bf x}) +1]
\nonumber \\
&& - [f_{{\bf K}}( {\bf x}) +1] [ f_{{\bf K}_2}( {\bf x}) +1]
 f_{{\bf K}_3}({\bf x}) f_{{\bf K}_4}( {\bf x}) \}
\nonumber\\
\end{eqnarray}
which now includes a streaming term. The collisional term in Eq.~(\ref{UU}) as
been approximated as s-wave scattering.

\subsubsection{Condensate master equation}
Finally, to illustrate the kinetics of Bose condensation we will derive in
Sec.~(\ref{Condeq})  a simple approximation of the QKME based on
the assumption that the density operator of the total system can be factorized
into a condensate density operator $\rho _0 (t)$ for the ${\bf K}=0$ Bose
condensate band, and an operator for the non--condensed modes ${\bf K} \ne 0$
which is assumed to be in thermal equilibrium.
We obtain the equation, which we shall call {\em the condensate master 
equation\/}:
\begin{eqnarray}\label{rho0}
\dot \rho _0( t)  &=&-{i\over\hbar}  \left[ \int d^3{\bf x}\,
\psi_0^{\dagger }( {\bf x}) \left( -{\hbar ^2\nabla ^2\over 2m}\right)
\psi _0({\bf x})
+
{1 \over 2 } u \int d^3{\bf x}
\, \psi _{{0}}^{\dagger }({\bf x})\psi _{{0}}^{\dagger}({\bf x})  \psi _{{ 0}}(
{\bf x}) \psi _{{0}}( {\bf x})\, , \, \rho_0 \right]\\
&&+
\left[(u g(0) \sum\limits_{{\bf K}\neq 0}\bar n_{{\bf K}})
 \int d^3{\bf x}\,\psi _0^{\dagger }( {\bf x}) \psi _0( {\bf x}) \, ,\,\rho_0
\right]
\nonumber \\
&&+\int d^3{\bf x}\int d^3{\bf x}'G^{(-) }( {\bf x-x}',T,\mu)
 \{ 2\psi _0( {\bf x}) \rho_0\psi _0^{\dagger }( {\bf x}') -\rho
_0\psi _0^{\dagger}( {\bf x}') \psi _0( {\bf x}) -\psi_0^{\dagger }(
{\bf x}') \psi _0( {\bf x})\rho _0\}
 \nonumber \\
 &&+\int d^3{\bf
x}\int d^3{\bf x}'G^{( +) }( {\bf x-x}',T,\mu)
\{ 2\psi_0^{\dagger }( {\bf x}) \rho _0\psi _0( {\bf x}') -\rho _0\psi _0({\bf
x}') \psi _0^{\dagger }( {\bf x}) -\psi_0( {\bf x}') \psi _0^{\dagger
}( {\bf x})\rho _0\} , \nonumber \\
\end{eqnarray}
Here $\psi_0 ({\bf x})$ refers to the atomic destruction operator for the ${\bf
K}=0$ band with the Bose condensate. The first line in Eq.~(\ref{rho0}) gives
the dynamic of the Bose condensate including the nonlinear interaction term
proportional to $\psi _{{0}}^{\dagger }\psi _{{0}}^{\dagger}  \psi _{{ 0}} \psi
_{{0}}$, and an interaction term of the atom in the Bose condensate with the
above condensate particles with thermal occupation numbers
$g(0) \sum\bar n_{{\bf K}\neq 0}$ (where $g(0)$ is a normalization factor to
be
defined in Eq.~(\ref{eq1.1902})).
For zero temperature, and when the field operators $\psi_0 ({\bf x})$ are
replaced by c--numbers, Eq.~(\ref{rho0}) is, of course, equivalent to the
Gross-Pitaevskii equation. For finite temperature, solutions of the
corresponding Hartree Fock equations have been discussed in references
\cite{Leggett,Siggia,Lovelace}.
The second and third line in Eq.~(\ref{rho0}) give dissipative loss and feeding
terms for the condensate due collisions with the non-condensed atoms.
The relevant collisions are of the form
${\bf K}_1, {\bf K}_2 \leftrightarrow {\bf K}_3, {\bf K}=0$, where atoms are
transferred to the Bose mode under
conservation of momentum and energy. In Eq.~(\ref{rho0}) these transition rates
are denoted to $G^{(\pm) }( {\bf x}-{\bf x}',T,\mu)$ (to be defined in
Sec.~\ref{eq4.12}); they involve the occupation numbers of the ${\bf K} \neq 0$
modes according to the given temperature and chemical potential


\subsubsection{Validity of the limiting cases}
These three cases are all very simplified, and more realistic applications of
the QKME will require much more care.  But each one them illustrates a
different aspect of the problem of quantum kinetics, which gives them great
value as an aid to intuition.

\section{Description of the system}\label{Sect. 6.2}
We consider a set of Spin-0 Bose particles, described by the
Hamiltonian
\begin{eqnarray}\label{eq1.1}
H=H_0+H_I + H_T,
\end{eqnarray}
in which
\begin{eqnarray}\label{eq1.2}
H_0=\int d^3{\bf x}\,\psi ^{\dagger }({\bf x})
\left(- {\hbar ^2\over2m}\nabla ^2\right) \psi ({\bf x})
\end{eqnarray}
and
\begin{eqnarray}\label{eq1.3}
H_I=\frac 12\int d^3{\bf x}\int d^3{\bf x}'\psi ^{\dagger }({\bf x})
\psi ^{
\dagger }( {\bf x}') u( {\bf x}-{\bf x}') \psi ( {\bf x}') \psi ( {\bf x}) .
\end{eqnarray}
Thus, the operators $ \psi ( {\bf x}) $ have the commutators
\begin{eqnarray}\label{eq1.4}
[ \psi ( {\bf x}) ,\,\psi ^{\dagger }( {\bf x}') ] =\delta ( {\bf
x}-{\bf x}'),
\end{eqnarray}
\begin{eqnarray}\label{eq1.5}
[ \psi ( {\bf x}) ,\psi ( {\bf x}')] =[ \psi ^{\dagger }( {\bf x})
,\psi ^{
\dagger}( {\bf x}') ] =0.
\end{eqnarray}
The potential function $ u( {\bf x}-{\bf x}')$ is a c-number and is as usual
not the true interatomic potential, but rather a short range
potential---approximately a delta function---which reproduces the correct
scattering length.  This enables the Born approximation to be applied, and thus
simplifies the mathematics considerably. \cite{Born approx}

The term $ H_T$ arises from a trapping potential, and is written as
\begin{eqnarray}\label{T1}
H_T =  \int d^3 {\bf x}\,V_T({\bf x})\psi^\dagger({\bf x})\psi({\bf x}).
\end{eqnarray}
In this paper we will restrict our considerations to the case of no trapping 
potential, so $ V_T$ will be set equal to zero.  The modifications necessary
if there is a nonzero trapping potential will be dealt with in a future 
publication.

Thus, this is the standard second quantized theory of an interacting Bose
gas of particles with mass $ m $.

\subsection{Possible phase space descriptions}\label{Sect. 6.2.1}
\subsubsection{The Wigner function}
The original work of Bose and Einstein \cite{Bose Einstein} was
conceived before the invention of quantum mechanics by Heisenberg
and Schr\"odinger, and was based
on a simple phase space description, in which both momentum and
coordinate space were discretized to give phase space cells of volume
$ h^3$.  The state of the system was then specified by the number of
quanta in each phase space cell.  The treatments in modern elementary
textbooks use essentially the same arguments, but use a single large
box in coordinate space, so that the discretizing comes about from the
discrete energy levels in this box.

We will want to develop a formulation which is as close to the
Boltzmann equation as is permitted by the quantum mechanical nature of
the problem. This requires a phase-space description of the field $
\psi ( {\bf x})$. Although it might seem that the Wigner function
provides the appropriate method, it is in fact impossible to write a
phase-space Wigner function for a multiparticle system which {\em
manifestly} exhibits the symmetry of the Bose wavefunction. A simple
demonstration of this problem can be made for a two particle system
whose Wigner function is obtained from the Fourier transforms with
respect to $ {\bf y}_1$, ${\bf y}_2$, of
\begin{eqnarray}\label{eq1.6}
&&{\rm Tr}\{ \psi ( {\bf x}_1+{\bf y}_1) \psi ( {\bf x}_2+{\bf y}_2)
\psi^{\dagger}({\bf x}_1-{\bf y}_1)\psi^{\dagger }({\bf x}_2-{\bf y}_2)\rho_0\}
\nonumber\\
&& \qquad\equiv \tilde{W}( {\bf x}_{1},{\bf y}_1,{\bf x}_{_2},{\bf y}_2) ,
\end{eqnarray}
where $ \rho_0 $ is the density operator of the vacuum. Thus, the
Wigner function is (where $ \,{\cal N}$ is an appropriate
normalization factor)
\begin{eqnarray}\label{eq1.7}
&& W( {\bf x}_{_1},{\bf k}_1,{\bf x}_2,{\bf k}_2) =
\nonumber\\
&&\qquad{\cal N}
\int e^{2i({\bf k}_1\cdot{\bf y}_1+{\bf k}_2\cdot{\bf y}_2)}
\tilde W({\bf x}_{_1},{\bf y}_1,{\bf x}_2,{\bf y}_2)d^3{\bf y_1}d^3{\bf y_2} .
\end{eqnarray}
The Wigner function has the symmetry $ 1\leftrightarrow 2 $ ; but this
is {\em not} the full Bose symmetry, which requires full symmetry
under the {\em independent} exchanges:
\begin{eqnarray}\label{eq1.8}
{\bf x}_1+{\bf y}_1/2 &\leftrightarrow &{\bf x}_2+{\bf y}_2/2,
\\
{\bf x}_1-{\bf y}_1/2 &\leftrightarrow &{\bf x}_2-{\bf y}_2/2\,.
\nonumber
\end{eqnarray}
The Wigner function symmetry $ 1\leftrightarrow 2 $ is given by the
{\em simultaneous} exchanges (\ref{eq1.8}), not the {\em independent}
exchanges. The problem obviously persists for all numbers of
particles. Where the Wigner function has been used in kinetic theory
\cite{Kirkwood} the Bose symmetry takes the form of a very
complicated integral operator relation, whose use is very impractical.
Although an exact evolution equation would preserve Bose symmetry, the absence 
of {\em manifest} Bose symmetry makes it extremely difficult to develop 
approximation methods which preserve Bose symmetry. 
In fact it is only in a very dilute gas limit, in which essentially
only the one-particle Wigner function occurs that the Wigner function has been 
used, and for this of course Bose symmetry is not an issue.

\subsubsection{Wavelet expansion of field operators}\label{Sect. 6.2.2}
The most natural way to give a phase space description of a Bose gas
is to follow the method of Bose and Einstein---that is, to divide
space into cells of volume $ \Delta V$, and for each cell to introduce
a set of basis wavefunctions which vanish outside the cell, but are
quantized within the cell.  Unfortunately, even from the point of view
of the non interacting Hamiltonian (\ref{eq1.2}), such wavefunctions
have infinite energy, arising from the sharp transition from inside to
outside the cell.  This yields a wavefunction which is not twice
differentiable, as required by (\ref{eq1.2}), so that the spread in
momentum $ {\bf p}$ is so large that the mean of $ {\bf p}^2 $ is
diverges.

To avoid this problem, the transition from inside the cell to outside
it must be made smoother, and this is a problem which is like that
which arises in the study of wavelets \cite{wavelets}.  We introduce a
set of {\em wavelet} functions (in one dimension)
\begin{eqnarray}\label{eq1.9}
v _{K}( x,r) &=&{1\over \sqrt{4\pi \Delta }}\int_{K-\Delta }^{K+\Delta
}e^{ik( x-r) }\,dk
\\ \label{eq1.10}
&\equiv &{e^{iK( x-r) }\over\sqrt{\pi \Delta }}\, {\sin\Delta (
x-r)\over x-r} .
\end{eqnarray}
If
\begin{eqnarray}\label{eq1.11}
r=n\pi /\Delta,\qquad n=0,\pm 1,\pm 2,...
\end{eqnarray}
these functions have the property of orthogonality:
\begin{eqnarray}\label{eq1.12}
\int_{-\infty}^{\infty} dx\,v_K^{*}( x,r) v_{K'}( x,r)
=\delta _{KK'}\delta _{rr'}
\end{eqnarray}
and they are also complete. From the definition (\ref{eq1.9}), it can
be seen that $ v_{K}( x,r) $ has momenta in the range $ (\hbar (
K-\Delta ) ,\hbar (K+\Delta ) ) $, and from of (\ref{eq1.10}), it can
be seen that wavefunction $ v_K( x,r) $ is localized to a certain
extent at the point $ x=r\equiv n\pi /\Delta $.

These wavefunctions correspond to a complete set of phase cells; we
see that if the uncertainties are defined as the intervals between the
discrete values of the momenta and position, then
\begin{eqnarray}\label{eq1.13}
\delta x=\pi /\Delta  \mbox{ and } \delta p=2\Delta \hbar
\Rightarrow  \delta x\,\delta p=h
\end{eqnarray}
However, the uncertainty as defined by a variance in $ x $ is clearly
infinite, reflecting the fact that the wavelet functions
(\ref{eq1.9}),(\ref{eq1.10}) are not well localized.  The quantities
$\delta x $, $ \delta p$ represent rather the spacing between the
phase cells---not the uncertainties in $x$ and $ p $.  In terms of the
variable $ x $, these wavelet functions represent the smoothest
possible functions we could choose, because by construction the
momentum spread is {\em bounded}, so that the wavelet functions are in
fact infinitely differentiable.

The field operators can now be expanded as (in an obvious
three-dimensional generalization)
\begin{eqnarray}\label{eq1.14}
\psi ( {\bf x}) =\sum_{{\bf K}}\sum_{{\bf r}}v_{{\bf K}}^{*}( {\bf x,r})
 a( {\bf r},{\bf K})
\end{eqnarray}
and the commutation relation for $ a( {\bf r},{\bf K}) $ is
\begin{eqnarray}\label{eq1.15}
[ a({\bf r},{\bf K}) ,a^{\dagger }({\bf r}',{\bf K}') ] = \delta
_{{\bf rr}'}\delta _{{\bf KK}'}\,.
\end{eqnarray}
The states of the Bose gas are now specified by the eigenvalues of the
number operators $ N({\bf r},{\bf K})$, which is a truly quantum
mechanical version of the original idea of Bose.

Notice that there is no need in this description for $ \Delta$ to be
unique---one can choose a different value of $ \Delta$  for each
$ {{\bf K}}$-band if one wishes, since momentum bands are orthogonal
independently of their size.

\subsubsection{Momentum resolved field operators}\label{Sect. 6.2.3}
For many parts of our discussion, it is convenient to resolve field
operators into only the different ranges of momentum; thus we write
\begin{eqnarray}\label{1.1501}
\psi({\bf x}) = \sum_{\bf K} e^{-i{\bf K\cdot x}}\psi _{{\bf K}}( {\bf x})
\end{eqnarray}
where
\begin{eqnarray}\label{eq1.16}
\psi _{{\bf K}}( {\bf x})  & = & e^{i{\bf K\cdot x}}\sum_{{\bf r}}v_{{\bf
K}}^{*}( {\bf x,r}) a( {\bf r},{\bf K}) \\
\label{eq1.17}
& =& e^{i{\bf K\cdot x}}\int D_{{\bf K}}( {\bf x}-{\bf x}') \psi (
{\bf x}')
\,d^3{\bf x}'
\end{eqnarray}
where
\begin{eqnarray}\label{eq1.18}
D_{{\bf K}}( {\bf x-x}')
=e^{-i{\bf K}.( {\bf x}-{\bf x}') }g({\bf x}-{\bf x}')
\end{eqnarray}
where
\begin{eqnarray}\label{eq1.1801}
g({\bf x}) ={1\over\pi ^3}\left[{{\sin \Delta x }}\over{{x}}\right]
\left[{{\sin \Delta  y }}\over{{y}}\right]
\left[{{\sin \Delta  z }}\over{{z}}\right] .
\end{eqnarray}
We also have the commutation relation
\begin{eqnarray} \label{eq1.1901}
[ \psi _{{\bf K}}( {\bf x}) ,\psi _{{\bf K}'}^{\dagger}( {\bf x}') ]
&=&\delta _{{\bf KK'}}g( {\bf x}-{\bf x}') .
\end{eqnarray}
The resolution into momentum bands thus yields a rather nonlocal
description.  Notice also that we can write
\begin{eqnarray}\label{eq1.1902}
g({\bf x}-{\bf x}') &=& {1\over
(2\pi)^3}\int_{-\Delta}^{\Delta}e^{i{\bf k}\cdot({\bf x} - {\bf x}')}
\, d^3 k \nonumber \\ &=&
\sum_{{\bf r}
}e^{-i{\bf K}\cdot ( {\bf x}-{\bf x}') }v_{{\bf K} }( {\bf x},{\bf r})
v_{{\bf K}}^{*}( {\bf x}^{\prime },{\bf r}) ,
\end{eqnarray}
which is an expression of the completeness of the wavelet functions within a 
$ K$-band, and is useful in some of the computations used in deriving the 
Uehling-Uhlenbeck equation in sect.\ref{Sect. 6.4}
\subsubsection{Free Hamiltonian in terms of momentum resolved field operators}
\label{Sect. 6.2.4}
Because the functions used in the expansion of $ e^{-i{\bf K\cdot
x}}\psi _{{\bf K}}( {\bf x}) $ are orthogonal to those used for $
e^{-i{\bf K}'\cdot {\bf x}}\psi _{{\bf K}'}( {\bf x}) $, it is obvious
that
\begin{eqnarray}\label{eq1.20}
\int d^3{\bf x\,}\psi _{{\bf K}}^{\dagger }( {\bf x}) \psi_{\bf K'}( {\bf x})
e^{-i( {\bf K}'-{\bf K}) \cdot {\bf x}} = 0
\quad\mbox{ if }{\bf K}\neq {\bf K}'.
\end{eqnarray}
Using this fact, it is straightforward to show that
\begin{mathletters}
\begin{eqnarray}\label{eq1.21a}
H_0 & =&\sum\limits_{{\bf K}}{{\hbar ^2{\bf K}^2}\over{2m}}\int d^3{\bf x}\,
\psi _{{\bf K}}^{\dagger }( {\bf x}) \psi _{{\bf K}}( {\bf x})
\\ \label{eq1.21b}
 &&+\sum\limits_{{\bf K}}\int d^3{\bf x\,}\hbar {\bf K} \cdot
{\bf j}_{{\bf K}}( {\bf x})
\\ \label{eq1.21c}
&&+\sum\limits_{{\bf K}}\int d^3{\bf x\,}\psi _{{\bf K}}^{\dagger}({\bf x})
\left(-{\hbar ^2\nabla ^2\over 2m}\right) \psi_{{\bf K}}( {\bf x})
\\ \nonumber
 &\equiv& H_a + H_b + H_c
\end{eqnarray}
\end{mathletters}
Here we have used a ``probability current'' for the momentum $ {\bf
K}$ defined in a way analogous to that normally employed
\begin{eqnarray}\label{1.210001}
{\bf j}_{{\bf K}}( {\bf x}) &\equiv&-{i\hbar\over 2m}\left\{
\psi_{\bf K}^{\dagger}({\bf r})\nabla\psi_{\bf K}({\bf r})
-\nabla\psi_{\bf K}^{\dagger}({\bf r})\psi_{\bf K}({\bf r})
\right \}.
\end{eqnarray}
The resolution into three parts has a simple interpretation. The
resolution into a full wavelet description shows that we can write
\begin{eqnarray}\label{eq1.22}
H_a=\sum\limits_{{\bf K,r}}\frac{\hbar ^2{\bf K}^2}{2m}a_{{\bf
K}}^{\dagger }( {\bf r}) a_{{\bf K}}( {\bf r})
\end{eqnarray}
corresponding to an energy $ {\hbar ^2{\bf K}^2}/{2m} $ for each
quantum at each phase space location $ {\bf K}, {\bf r}$.  This is the
Hamiltonian corresponding to the original ideas of Bose and Einstein.

The term $ H_b $ corresponds to transport, as shown by computing the
commutator\begin{eqnarray}\label{eq1.2201}
\big [H_b, \psi_{\bf K}({\bf x}) \big] &=&
i \hbar {\bf v}_{\bf K}\cdot\nabla\psi_{\bf K}({\bf x})
\end{eqnarray}
where $ {\bf v}_{\bf K} = \hbar{\bf K}/m$.  This is obviously
transport corresponding to the velocity appropriate to the central
momentum of the momentum band $ {\bf K}$.

Finally, the last term $ H_c$ corresponds to wavepacket spreading
through
\begin{eqnarray}\label{eq1.2202}
\big [H_c, \psi_{\bf K}({\bf x}) \big] &=&-{\hbar^2\over 2m} \nabla^2 \psi_{\bf
K}({\bf x})
\end{eqnarray}
Thus the wavelet description naturally gives the separation into the
three parts fundamental to a phase space description of the processes.

\subsubsection{Interaction Hamiltonian in terms of the momentum resolved field
operators}\label{Sect. 6.2.5} It is trivial that the interaction part,
$ H_I $, can be written
\begin{eqnarray}\label{eq1.23}
H_I=\frac 12\sum\limits_{{{\bf K}_1,{\bf K}_2 ,{\bf K}_3 ,{\bf K}_4}}
U( {\bf K}_1,{\bf K}_2,{\bf K}_3,{\bf K}_4) ,
\end{eqnarray}
in which
\begin{eqnarray}\label{eq1.24}
&&U( {\bf K}_1,{\bf K}_2,{\bf K}_3,{\bf K}_4) =
\nonumber \\
&&\qquad\int d^3{\bf x}\int d^3{\bf x}'
\, e^{({i{\bf K}_1{\bf \cdot x}+i{\bf K}_2{\bf \cdot x}'-i{\bf K}_3{\bf
\cdot x}-i{\bf K}_4{\bf \cdot x}'})}
\nonumber \\
&&\qquad \times\psi _{{\bf K}_1}^{\dagger }({\bf x})\psi _{{\bf
K}_2}^{\dagger}({\bf x}') u({\bf x}-{\bf x}') \psi _{{\bf K}_3}( {\bf
x}') \psi _{{\bf K}_4}( {\bf x}).
\end{eqnarray}
Furthermore, it is easy to see that
\begin{eqnarray}
U( {\bf K}_1,{\bf K}_2,{\bf K}_3,{\bf K}_4) =0
\end{eqnarray}
unless
\begin{eqnarray}\label{eq1.25}
{\bf K}_1+{\bf K}_2={\bf K}_3+{\bf K}_4+ O( {\Delta } )
\end{eqnarray}
i.e., the quasimomentum $ {\bf K}_1 $ is conserved to within something
not much bigger than $ \Delta $.

\section{Treatment of collisions}\label{Sect. 6.3}

\subsection{Eigenstates of  $ H_a $  and its corresponding Liouvillian}
\label{Sect. 6.3.1}
Suppose we use the wavelet basis; then we can label the eigenstates of
$ H_a\, $ (as given in (\ref{eq1.21a})) as $ | {\bf n} \rangle $ where
$ {\bf n} $ is a vector of elements $ n_i $, the number of quanta
belonging to the phase space wavelet $ i $, which has quantum number $
( {\bf K}_i,{\bf r}_i) $. The eigenstates of the corresponding
Liouvillian are given by the outer product $ | {\bf n}\rangle\langle
{\bf m}| $, and the eigenvalue is only zero if the energy of $ | {\bf
n}\rangle $ is the same as that of $ | {\bf m}\rangle $.

\subsection{Projectors}\label{Sect. 6.3.2}
We will define a projector onto all eigenstates with the same number
of quanta in each $ {\bf K} $, irrespective of $ {\bf r} $; thus
we define
\begin{eqnarray}\label{eq2.1}
p_{\bf N} |{\bf n\rangle } & =& |{\bf n\rangle }\quad\mbox{ if }
\sum_{\bf r} n({\bf K},{\bf r}) =N( {\bf K})
\mbox{ and } {\bf K}\ne 0;
\nonumber\\
& =& 0 \qquad\mbox{ otherwise.}
\end{eqnarray}
This kind of projector identifies all configurations with the same
distribution in $ {\bf K} $, but ignores $ {\bf K}=0\, $ and leaves
undisturbed the distribution over $ {\bf r} $.

Suppose now that $ \rho $ is the density operator for the system of
particles; we define a projector on $\rho $ by
\begin{eqnarray}\label{eq2.2}
{\cal P}_{\bf N} \rho =p_ {\bf N}\, \rho\, p_ {\bf N} \equiv v _ {\bf
N} .
\end{eqnarray}
Thus the projected $ \rho $ is an operator in position space, but acts
only with a space that has the configuration $ {\bf N} $ of particles
over the momentum space cells---except any configuration of $ {\bf
K}=0 $ particles is permitted (we call this $ {\bf K} $ diagonal). It
follows also that, for any $ \rho$
\begin{mathletters}
\begin{eqnarray}\label{eq2.3}
{\cal P}_{\bf N} H_b\rho= H_b{\cal P}_{\bf N}\rho\\
\label{eq2.301}
{\cal P}_{\bf N} H_c\rho= H_c{\cal P}_{\bf N}\rho
\end{eqnarray}
\end{mathletters}
and that for $ H_{a} $ there is the stronger result
\begin{eqnarray}\label{eq2.4}
\big[ H_a, {\cal P}_ {\bf N} \rho \big] =0
\end{eqnarray}
which is true by construction.

The thrust of this paper is now to develop a closed equation of motion for the 
projected parts of $ \rho(t)$, namely $ v_{\bf N}(t)$, and to show that we can 
use these parts to represent the physics in which we are interested.  Thus we 
will be making the approximation
\begin{eqnarray}\label{thru1}
\rho(t) \approx \sum_{\bf N} v_{\bf N}(t),
\end{eqnarray}
which will require the assumption that we can neglect the remainder: thus
\begin{eqnarray}\label{thru2}
w(t) &\equiv& \left(1- \sum_{\bf N}{\cal P}_{\bf N}\right)\rho(t)
\nonumber \\ &\approx & 0.
\end{eqnarray}
A density operator of the form (\ref{thru1}) will be called 
{\em $ K$-diagonal}. For such $ K$-diagonal density operators, the algebra in 
deriving the equations of motion is considerably simplified by results like
\begin{eqnarray}\label{thru3}
&&{\cal P}_{\bf M}\{\psi^\dagger_{\bf K}({\bf x})\psi_{\bf K'}({\bf x}')
v_{\bf N}\} 
=0
\nonumber\\
&&\qquad\mbox{ unless $ {\bf K}= {\bf K}'$ and $ {\bf M} = {\bf N}$.}
\end{eqnarray}
Notice however that in this equation, we do not require that 
$ {\bf x} = {\bf x}'$; thus the designation $ K$-diagonal is appropriate.
Another kind of identity is 
\begin{eqnarray}\label{thru4}
&&{\cal P}_{\bf M}\{\psi^\dagger_{\bf K}({\bf x})
v_{\bf N} \psi_{\bf K'}({\bf x}')\}
=0
\nonumber \\
&&\qquad\mbox{ unless $ {\bf K}= {\bf K}'$ and $ {\bf M} = {\bf N + b }$.}
\end{eqnarray}
where $ {\bf b}$ is a vector in the same space as $ {\bf M}$ and $ {\bf N}$,
and whose only non-zero component is $ b({\bf K})=1$, so that $ {\bf M}$ and 
$ {\bf N}$ are identical apart from the change 
$M({\bf K})=N({\bf K})+1$.

There are many other relations like these, but all are essentially of the same 
kind, that the $ K$-diagonal property is preserved only by matching a creation 
field operator with a destruction field operator with the same $ {\bf K}$, 
though not necessarily the same $ {\bf x}$, either on the same or the other 
side of the density operator.  If the matching is done on the same side, the 
configuration $ {\bf N}$ is preserved, if the matching is done on opposite 
sides the configuration changes.
\subsection{Derivation of the quantum kinetic master equation}
\label{Sect. 6.3.3}
The equation of motion for the density operator is
\begin{eqnarray}\label{eq2.5}
\dot \rho &=& -{ i\over\hbar} [ H_a+H_b+H_c,\rho]-{i\over\hbar}[H_I,\rho] \\
\label{eq2.6}
&=&{\cal L}_a\rho +{\cal L}_b\rho +{\cal L}_c\rho +{\cal L}_2\rho .
\end{eqnarray}
We define the Laplace transform
\begin{eqnarray}\label{eq2.7}
\tilde \rho ( s) =\int_{_0}^\infty e^{-st}\rho ( t) dt
\end{eqnarray}
and correspondingly
\begin{eqnarray}\label{eq2.8}
\tilde{v}_{{\bf N}}( s) & = & {\cal P}_ {\bf N} \tilde\rho(s)  \\
\label{eq2.9}
\tilde{w}(s) & = & {\cal Q}\tilde\rho( s)
\equiv\left[ 1-{\sum_N} {\cal P} _ {\bf N} \right] \tilde{\rho }(s) ,
\end{eqnarray}
so that
\begin{eqnarray}\label{eq2.10}
&& s\tilde v _{{\bf N}}( s) -v _{{\bf N}}(0)
\nonumber \\
&&\quad
=( {\cal L}_b+{\cal L}_c)
\tilde v _{{\bf N}}( s) +{\cal P}_ {\bf N} {\cal L}_2
\left\{ \sum_{\bf M}\tilde v_{{\bf M}}( s) +\tilde{w}( s) \right\}
\end{eqnarray}
and
\begin{eqnarray}\label{eq2.11}
&& s\tilde{w}(s)-w(0)
\nonumber \\
&&\quad
=\left\{{\cal L}_a+{\cal L}_b+{\cal L}_c\right\}\tilde{w}(s)
+{\cal QL}_2\left\{\sum\limits_{{\bf M}}\tilde v_{{\bf M}}(s)+\tilde{w}(s)
\right\}.
\end{eqnarray}
We assume $ w( 0) =0, $ that is the state that is already $K$-diagonal.
This assumption cannot be made without some justification, 
which we shall postpone to Sect.~\ref{justification}.  Obviously we can 
choose any initial condition we wish; the justification required is that when 
we solve the equations of motion it will be possible to show that $ w(t)$ 
rapidly becomes negligible, and 
negligible at all times, so that any time can be chosen as an initial time.

From this we can then readily derive
\begin{eqnarray}\label{eq2.12}
&& s\tilde v _{{\bf N}}(s)-v_{{\bf N}}(0) =( {\cal L}_b+{\cal L}_c)
\tilde v_{{\bf N}}(s)+{\cal P}_{\bf N}{\cal L}_2\sum\limits_{\bf M}\tilde v_{
\bf M}( s)
\nonumber \\
&&\quad
+ {\cal P}_{\bf N} {\cal L}_2[ s-{\cal L}_a-{\cal L}_b-{\cal
L}_c-{\cal QL}_2] ^{-1} {\cal QL}_2\sum\limits_{{\bf M}}\tilde v_{{\bf
M}}( s) .
\end{eqnarray}
This equation is exact; the physical content comes into the
assumptions concerning what parts can be considered to be negligible.
We invert the Laplace transform in (\ref{eq2.12}), and consider the
parts separately.

\subsubsection{Streaming and Quantum Terms}
These are the names we use for $ {\cal L}_b $ and $ {\cal L}_{c\,} $
as defined in (\ref{eq2.6}), (\ref{eq1.21b}, \ref{eq1.21c}). They keep the same
form with no approximations.
\subsubsection{The Forward Scattering Terms}
The term $ {\cal P}( {\bf N}) {\cal L}_2\sum\limits_{{\bf M}}\tilde v
_{{\bf M}}( s) \, $ can be simplified by noting that $ v _{{\bf M}}(
s) $ is $K$-diagonal, and that $ {\cal P} ( {\bf N}) $ projects onto a
$K$-diagonal density operator.  Since $ {\cal L}_2 $ is a commutator;
i.e.,
\begin{eqnarray}\label{eq2.13}
{\cal L}_2\sum\limits_{{\bf M}}v _{{\bf M}}( s) =-\frac i
\hbar \sum\limits_{{\bf M}}[ H_I,v _{{\bf M}}( s)]
\end{eqnarray}
this can only yield a $ K $-diagonal term from those parts of $ H_I $
which do not change the $ {\bf K} $-distribution; that is, the terms
in $ H_I $ given by $ H_F $, which we define as
\begin{eqnarray}\label{eq2.14}
H_F &\equiv &
 \frac 12\sum\limits_{{\bf K}_1 \ne {\bf K}_2}
U( {\bf K}_1,{\bf K}_2,{\bf K}_1,{\bf K}_2)
\nonumber\\
&& + \frac 12\sum\limits_{{\bf K}_1 \ne {\bf K}_2}
U( {\bf K}_1,{\bf K}_2,{\bf K}_2,{\bf K}_1)
\nonumber\\
&& +
 \frac 12\sum\limits_{{\bf K}_1}U( {\bf K}_1,{\bf K }_1,{\bf K}_1,{\bf K}_1) .
\end{eqnarray}
This thus gives only transitions between the same $ {\bf K} $
distribution, and this can be regarded as forward scattering.  These forward
scattering terms give rise to the so-called {\em mean field effects}, that is
the average effect on the motion of one particle of the interaction with all
other particles.
\subsubsection{The Collision Terms}
We now introduce a Born approximation, which amounts to the neglect of
$ {\cal L}_b+{\cal L}_c+{\cal QL}_2 $ in the $ [ \quad] ^{-1} $ term in
(\ref{eq2.12}), leaving a contribution to $ \dot v_{{\bf N}}(t) $
given by
\begin{eqnarray}\label{eq2.15}
\sum\limits_{{\bf M}}{\cal P}_{{\bf N}}{\cal L}_2\int_{_0}^td\tau
 \exp\{ {\cal L}_a\tau \} {\cal QL}_2 v_{{\bf M}}(t-\tau).
\end{eqnarray}
Notice however that $ {\cal QL}_2$ contains terms which depend on
$\psi_0 $, $ \psi^\dagger_0$, and when there is significant condensation these
terms can become very large.  Thus this approximation is valid only in the case
of {\em weak condensation}.  This means that we approximate the free evolution
operator by discretizing the eigenvalues to those corresponding to the centers
of the momentum bands.  This leads to an approximate measure of the validity of
the neglect of the terms  $ {\cal QL}_2 $ in comparison to the term retained,
${\cal L}_a $, which is given by requiring that the modification by the
presence of any condensate to the excitation spectrum should be negligible for
energy greater than $ \hbar^2\Delta^2 / 2 m$.  Using the Bogoliubov theory,
this leads to the conclusion that we must have the condition on the cell size
\begin{eqnarray}\label{condition1}
l_c = {\pi\over\Delta} &\gg & {{\pi\hbar\over \sqrt{2 m \rho u}}}
\nonumber \\
&=&  \sqrt{\pi\over 8 a\rho}
\end{eqnarray}
where $ \rho$ is the density of particles.
We now want to make a Markov approximation, which involves two steps.
Firstly we neglect the $ \tau $ dependence in the last term, and secondly we
let the upper limit in the integral become infinite. For this approximation to
be valid there must be a smooth distribution over the available energy states,
so that when we sum over all $ {\bf M}$, the resultant range of $ \tau$
in which the integrand is non-zero  is very much smaller then the
characteristic  time over which $ v_{\bf M}(t)$ evolves. This is a requirement
on the kind of $ v_{{\bf M}}(t)$ being considered.

Now define the vector in the space of
$ {\bf M},{\bf N}, $ called $ {\bf e}(1234) $
which can be written
\begin{eqnarray}\label{eq2.17}
&&{\bf e}(1234)_{{\bf K}_1}={\bf e}(1234) _{{\bf K}_2} = -{\bf e}( 1234)
_{{\bf K}_3}=-{\bf e}( 1234) _{{\bf K}_4}=1.
\nonumber \\&&
\end{eqnarray}
and all other components are zero, and let
\begin{eqnarray}\label{eq2.18}
\hbar \omega _{{\bf K}}=\frac{\hbar^2 {\bf K}^{2}}{2m}
\end{eqnarray}
We also introduce the notation
\begin{eqnarray}\label{eq2.2101}
\Delta\omega_{\bf e} = \omega_4+\omega _3-\omega _1-\omega _2,
\end{eqnarray}

Before we examine the range of validity of them Markov conditions, we can see 
that using them (\ref{eq2.15}) now becomes
\begin{eqnarray}\label{eq2.16}
&&  -{1\over4\hbar ^2}{\cal P}_{\bf N} \sum\limits_{{1234 \atop
1'2'3'4'}} \bigg\{\bigg[ U( 1'2'3'4') ,
\nonumber\\
&&\quad\int_{_0}^\infty d\tau\exp (i\Delta\omega_{\bf e}\tau)
 {\cal Q}\big[ U( 4321)  ,\sum\limits_{_{{\bf M}}}v _{{\bf M}}( t)
\big]\bigg] \bigg\}.\nonumber \\
\end{eqnarray}
Since the state $ v_{{\bf M}}(t) $ is $K$-diagonal, and we project
onto $K$-diagonal states, we must have
\begin{eqnarray}
1=1',2=2' &\mbox{ or }&1=2', 2=1' \nonumber\\ &\mbox{and}& \nonumber\\
3=3',4=4' &\mbox{ or }&3=4', 4=3' \nonumber
\end{eqnarray}
but since the operator $ U(1234) $ is symmetric in $ (12) $ and $ (34)
$ this simply gives $ 4 $ times the result obtained from setting
$ (1234) =( 1'2'3'4') $.

Notice that no ``forward scattering terms'' arise in (\ref{eq2.16}).
These have already been explicitly separated from the scattering terms and 
included in the term $ H_F$ defined in (\ref{eq2.14}), and are eliminated from 
(\ref{eq2.16}) by the $ {\cal Q}$ projection operator.

We can now carry out the time integral
so that (\ref{eq2.16})becomes
\begin{eqnarray}\label{eq2.19}
&& - \lim\limits_{\epsilon \to 0}
\sum\limits_{{\bf e}}\left\{
 U_{\bf e}U^{\dagger }_{\bf e} v_{{\bf N}}+
v_{{\bf N}} U^{\dagger }_{\bf e} U_{\bf e}
 -U_{\bf e} v_{{\bf N- e}}U^{\dagger }_{\bf e}  -U^{\dagger }_{\bf e}
 v_{{\bf N+e} }U_{\bf e}  \over \hbar ^2( \epsilon +i\Delta\omega_{\bf e} )
\right\}.
\nonumber\\
\end{eqnarray}
Here we have used $ U({1234})^{\dagger }=U( 4321)$. We now use
\begin{eqnarray}\label{eq2.1901}
{1 \over z+i\epsilon }\to {\rm P}{1\over z}-i\pi\delta(z)
\end{eqnarray}
and rearrange, so that (\ref{eq2.19}) becomes
\begin{eqnarray}\label{eq2.20}
&&{ -{i\over\hbar ^2}\sum\limits_{{{\bf e}}}{\rm
P}{1\over\Delta\omega_{\bf e}} [ U_{\bf e}U^{\dagger }_{\bf e} ,v_{{\bf N}}]}
\nonumber\\ &&{ +{\frac \pi {\hbar
^2}}\sum\limits_{{\bf e}}{\delta (\Delta\omega_{\bf e}) }} \{ 2U_{\bf e}
 v _{{\bf N}-{\bf e}}(t) U^{\dagger }_{\bf e}
- U^{\dagger}_{\bf e} U_{\bf e} v _{{\bf N}}
-v_{\bf N} U^{\dagger}_{\bf e}U_{\bf e} \} .
\end{eqnarray}

As is usual, we find a {\em level shift} term, which is a commutator,
and a purely dissipative term. Adding together (\ref{eq2.20}),
(\ref{eq2.13}),
(\ref{eq2.6}), we get the full Quantum Kinetic Master Equation (QKME)
\begin{mathletters}
\begin{eqnarray}\label{eq2.21a} 
 \dot v_{{\bf N}}( t) &=& -{ i\over\hbar}\left[
 \sum\limits_{{\bf K}}\int d^3{
\bf x\,}{\left(\hbar {\bf K} \over 2m\right)}\cdot {\bf j}_{{\bf K}}( {\bf x})
,v _{{\bf N}}(t) \right]
\\ \label{eq2.21b}
 &&-{ i\over\hbar} \left[
\sum\limits_{{\bf K}}\int d^3{\bf x\,} \psi _{{\bf K}}^{\dagger }({\bf
x})\left(-{\hbar ^2\nabla^2\over 2m}\right)
\psi_{{\bf K}}( {\bf x}) ,v_{{\bf N}}(t) \right]
\\ \label{eq2.21c}
&&{ - {i\over 2\hbar} \left[
\sum\limits_{{\bf K}_1 \ne {\bf K}_2}
\big(U( {\bf K}_1,{\bf K}_2,{\bf K}_1,{\bf K}_2) +
U( {\bf K}_1,{\bf K}_2,{\bf K}_2,{\bf K}_1)\big)
 +\sum\limits_{{\bf K}_1}U( {\bf K}_1,{\bf K }_1,{\bf K}_1,{\bf K}_1)
   	,v_{{\bf N}}(t)\right] }
\\  \label{eq2.21d}
&& {- {i\over\hbar ^2}\sum\limits_{{\bf e}}}{\rm P}
{1\over\Delta\omega({\bf e})} \left[U({\bf e})
   	U^{\dagger }({\bf e}),v_{\bf N}(t) \right]
\\  \label{eq2.21e}
&&{ +{\pi \over\hbar
^2}{\sum\limits_{{\bf e}}}\delta (\Delta\omega({\bf e}))} \{ 2U({\bf
e})v_{{\bf N-e}}(t) U^{\dagger}({\bf e})
-U^{\dagger}({\bf e})U({\bf e})v_{{\bf N}}(t)-v_{{\bf N}}(t)
U^{\dagger}({\bf e}) U({\bf e})\}.
\end{eqnarray}
\end{mathletters}


\subsection{Approximations}
\subsubsection{Weak condensation condition}
Approximations only occur in the derivation of the
collision and level shift terms. The major approximation is the
neglect of ${\cal L}_b+{\cal L}_c+ {\cal QL}_2 $ compared to $ {\cal
L}_a $. We choose $ \Delta $ so that eigenvalues of $ {\cal L}_a\, $
(apart from zero eigenvalues) are substantially larger than typical
measures of the size of $ {\cal Q}{\cal L}_2$ on the particular states
involved.
Since the projectors $ {\cal P}_{\bf N}$ do not affect the
${\bf K} =0$ composition of the states, the operator $ {\cal Q}{\cal L}_2 $
includes the term
$ (u/2)\int d^3{\bf x}\,\psi_0^\dagger\psi_0^\dagger\psi_0\psi_0$, which
can become very large when there is significant condensation, hence this
approximation will only be valid in the case of weak condensation, the
condition for which is given by (\ref{condition1}); that is
\begin{eqnarray}\label{condition1a}
l_c = {\pi\over\Delta} &\gg & {{\pi\hbar\over\sqrt{ 2 m \rho u}}}
\nonumber \\
&=&  \sqrt{\pi\over 8 a\rho}
\end{eqnarray}
We will deal with the issues of strong condensation in a subsequent paper.

\subsubsection{The Markov Approximation}
The Markov approximation is really a kind of perturbative
result. The change from the integral
 $ \int_0^t d\tau \to \int_0^\infty d\tau  $  requires the assumption that the
terms that turn up have a broad spectrum of frequencies which are almost
continuous.  The frequencies which do turn up are of the form
$ \omega_{\bf e}$, and the range of these is of the order $ kT/\hbar$, while
a typical separation between the frequencies corresponding to adjacent
transitions
with a typical magnitude of momentum $  \sqrt{2 m k T}$ is of order of
magnitude $\Delta \sqrt{2  k T/m} $, which is the inverse of the time required
for a particle to traverse a cell at a typical thermal velocity.

The condition that we can regard the frequency spectrum as almost continuous
is that the separation between the frequencies is very much less than their
range, that is, the density of states is sufficiently high. Quantitatively, 
this leads to
\begin{eqnarray}\label{condition2}
l_c \gg \lambda_T \equiv h/ \sqrt{2mkT}.
\end{eqnarray}

If the density operators $ v_{\bf N}$ are such that the contributions  from 
sums over the different $ {\bf e}$ are all smoothly varying functions of the
contributing frequencies, then the time dependence given by summing up over all
transitions in (\ref{eq2.16}) will drop of over a characteristic time given by
the inverse of the bandwidth, i.e, $ \hbar/ kT $, and this time must be very
much less than the typical time scale of evolution of the the distribution,
which is of the order of magnitude of the time between collisions.  This leads
to the requirement on the mean free path $ \lambda_{\rm mfp}$
\begin{eqnarray}\label{condition3}
\lambda_{\rm mfp} \gg \lambda_T.
\end{eqnarray}

Finally, the QKME itself gives a level broadening, which is implicitly small
compared to the distance between levels.  The condition for this must be that
there is little likelihood of a collision as a particle traverses a single
cell, which means of course that
\begin{eqnarray}\label{condition4}
\lambda_{\rm mfp} \gg l_c
\end{eqnarray}.

All these conditions can be consistently satisfied.

Coming back now to the quantum kinetic master equation we see that the
summations $ \sum_{{\bf e}} $ are over the discrete momentum ranges $
{\bf K}_{1},{\bf K}_{2},{\bf K}_{3},{\bf K}_{4} $, and the energy
conservation delta function is sharp.
This sharpness is really an artifact which arises from the extension of the
upper limit of the time integral to infinity.  This integral can be cut off at
any time very much larger than the characteristic decay time of the kernel,
$ \hbar/ kT $.  This means that the delta function can be taken as being
broadened correspondingly, and it is easy to see that this broadening can
encompass a significant number of momentum bands. Nevertheless, this broadened
delta function can be written as an exact delta function if the remainder of
the integrand is smooth, and the sums written as integrals using
the continuum approximation
$\sum\limits_{{\bf K}}\to \int d^3{\bf K} $.

However, we have already assumed this smoothness in establishing the validity
of the Markov approximation so this procedure is consistent with the Markov
approximation.

\subsubsection{The neglect of $ w(0)$}\label{justification}
We will assume $ w(0)=0$, and then make an estimate of the size of $ w(t)$ and 
find under what conditions that it is negligible for all future times.  
From (\ref{eq2.11}) it follows, on inverting Laplace transforms, that
\begin{eqnarray}\label{just1}
w(t) &=& \int_0^t\exp\left\{({\cal L}_a + {\cal L}_b + {\cal L}_c 
         +{\cal Q}{\cal L}_2)(t-t')\right\}
\nonumber\\
&&\times\sum_{\bf M}v_{\bf M}(t')dt'.
\end{eqnarray}
We can now approximate this by keeping $ {{\cal L}}_a$, as in the derivation of 
the QKME, and replacing the effect of the all the other terms, including the 
interactions, by a phenomenological decay term, so that we write approximately
\begin{eqnarray}\label{just2}
w(t) &\approx& \int_0^t  e^{\left\{({\cal L}_a -\gamma)(t-t')\right\}} 
            \sum_{\bf M}v_{\bf M}(t')dt'
\\
&=& -{i\over\hbar}\sum_{\bf e}\sum_{\bf M}
    e^{(i\Delta\omega_{\bf e}-\gamma)(t-t')}[U({\bf e}),v_{\bf M}(t')]dt'.
\end{eqnarray}
As in the QKME derivation, we make the replacement 
 $v_{\bf M}(t') \to  v_{\bf M}(t) $, and now we can do the integration over 
$ t'$ to get approximately
\begin{eqnarray}\label{just3}
w(t) &\approx& 
 -{i\over\hbar}\sum_{\bf e}\sum_{\bf M}
   {1\over \gamma -i\Delta\omega_{\bf e}}[U({\bf e}),v_{\bf M}(t)].
\end{eqnarray}
For any given $ {\bf M}$ and $ {\bf e}$ we get a specific non $ K$-diagonal 
term.  The largest of these will occur when $\Delta\omega_{\bf e} =0$.  To 
estimate 
their size, we go to the definition (\ref{eq1.24}) of $ U({\bf e}) $.  
Substituting the expansion on wavelet functions into this equation, and 
assuming 
the approximation $ u({\bf x}) = u\delta({\bf x})$, we find that the largest 
terms that occur are of the form
\begin{eqnarray}\label{just4}
&&\int\,d^3{\bf x}/,
v_{\bf K_1}({\bf x},{\bf r})e^{i{\bf K_1}\cdot{\bf x}}
v_{\bf K_2}({\bf x},{\bf r}))e^{i{\bf K_2}\cdot{\bf x}}
\nonumber \\
&&\times
v^*_{\bf K_3}({\bf x},{\bf r}))e^{-i{\bf K_3}\cdot{\bf x}}
v^*_{\bf K_4}({\bf x},{\bf r}))e^{-i{\bf K_4}\cdot{\bf x}}
\nonumber\\
&& \times
a^{\dagger}({\bf K_1},{\bf r})
a^{\dagger}({\bf K_2},{\bf r})
a({\bf K_3},{\bf r})
a({\bf K_4},{\bf r})
\end{eqnarray}
in which all wavelet functions refer to the same cell.  Using the definitions 
of the wavelet functions, the coefficient of the operator part in this equation 
can be estimated to be of order of magnitude  $1/l_c^{3} $  The effect of the 
creation and destruction operators is of order of magnitude $ 1$, so we can now 
estimate the size of the term in $ w(t)$ compared to those in $ v_{\bf M}(t)$ 
to be of order of magnitude given by
\begin{eqnarray}\label{just5}
&&w(t)\approx
{a\hbar\over \gamma l_c^3}
\nonumber\\ \quad
&&\times\sum_{\bf e,M,r}
[a^{\dagger}({\bf K_1},{\bf r})
a^{\dagger}({\bf K_2},{\bf r})
a({\bf K_3},{\bf r})
a({\bf K_4},{\bf r}),v_{\bf M}(t)].
\end{eqnarray}
The size of the initial coefficient can be made clearest by taking 
$ \gamma$ to be given by $ v_T/\lambda_{\rm mfp}$, in terms of which the 
coefficient can be simplified, leading to the condition for the validity of the 
neglect of non-$K$-diagonal terms:
\begin{eqnarray}\label{just6}
{a\lambda_{\rm mfp}\lambda_T\over l_c^3} \ll 1.
\end{eqnarray}
This condition can be satisfied simultaneously with the two other conditions
(\ref{condition2}) and (\ref{condition3}).  For example, for sodium at 
$ T=2\mu{\rm K}$, the scattering length is $ a= 4.9{\rm nm}$, leading to
\begin{eqnarray}\label{just7}
\lambda_{\rm mfp}  	&=& 0.42{\rm m}\\
\lambda_T  	   	   	&=& 4.8\times10^{-7}{\rm m}\\
 a 	   	   	   	   	&=& 4.9\times 10^{-9}{\rm m}\\
\left(a\lambda_{\rm mfp}\lambda_T\right)^{1/3}
   	   	   	   	   	&\approx&  10^{-5}{\rm m}
\end{eqnarray}
Thus if $ l_c$ is chosen in the range somewhat greater than $  10^{-5}{\rm m}$
this kind of treatment will be valid.

The condition (\ref{condition1}) is best viewed as a condition on the 
condensate 
density $ \rho$; insertion of the sodium data into it gives the requirement 
that $ \rho\ll 10^{18}{\rm m}^{-3}$, which can be satisfied, but will not 
always be satisfied in current experiments.   In fact this density corresponds 
to about one atom per cube of volume $ \lambda_T^3$ using the above data, 
which  indicates that for an accurate treatment of condensation, the weak 
condensation assumption will  be invalid in this case.

It should be borne in mind that all of the above has been in the absence of a 
trapping potential---the question will be revisited in our forthcoming paper
dealing with strong condensation.

\subsubsection{Born Approximation}
The derivation of the collision terms relies on the Born
approximation, and the collision rates appearing in the QKME
(\ref{QKME}) are proportional to the Born scattering cross section.
This implies that the potential should be weak, i.e.~a small
perturbation on the energy scale associated with the momentum coarse
graining. In derivations of the quantum Boltzmann equation it is shown
\cite{Born approx}
that in a binary collision approximation, which assumes a low density
but not necessarily a weak potential ($\rho a^3 \ll 1$), the Born
amplitude should be replaced by the two particle $T$-matrix element
(proportional to the scattering amplitude $a$ in low energy s-wave scattering).
Essentially the same situation should pertain in this case and this
means that we replace the exact interatomic potential with one for which the
Born approximation {\em is} valid, and whose scattering length is the same as
that of the exact potential.  However, a proof of this assertion in our
formulation is at present lacking.
\subsubsection{Level shifts}
What are normally called ``collisional level shifts'' appear in our formalism 
in two guises.  There are the forward scattering terms which generate $ H_F$, 
as defined in (\ref{eq2.14}), and there are the higher order terms which from 
the derivation of the collisional terms in (\ref{eq2.21d}).  In the weak 
condensation situation which we want to treat in this paper, we expect the 
density to be so low that these can be ignored, but in our forthcoming work it 
will not be possible to do so without more careful justification.

\subsection{Stationary Solution}
Let us assume the level shift term is small---negligible in fact. Then it is
clear, since none of the other
terms can actually change the eigenvalue of $ H_a $, (the total coarse
grained kinetic energy) or the total number of particles, that any
function of
\begin{eqnarray}  \label{eq2.22}
H_a &=&\sum\limits_{{\bf r},{\bf K}}{\hbar ^2{\bf K}^2\over 2m}
a_{{\bf K}}^{\dagger }({\bf r}) a_{{\bf K}}( {\bf r})
\end{eqnarray}
and
\begin{eqnarray}\label{eq2.2201}
N      &=&   \sum\limits_{{\bf r},{\bf K}}
a_{{\bf K}}^{\dagger }({\bf r})    a_{{\bf K}}({\bf r})
\end{eqnarray}
will provide a stationary solution---clearly microcanonical,
canonical, and grand canonical versions will exist.

Thus the result is the statistical mechanical result---but coarse
grained in momentum space---the size $ \Delta $ of the momentum coarse
graining to be such that $ H_I $ is negligible compared with it on the
states of interest.
\subsubsection{Correlation Functions in the Stationary Solution}
Since the stationary solution can be written as a function of $ H_a$
and $ N $, the correlation functions of the wavelet creation and
destruction operators can be written as
\begin{eqnarray} \label{eq2.23}
\langle a_{{\bf K}}^{\dagger }( {\bf r})\, a_{{\bf K}'}( {\bf r}') \rangle =
\bar{N}_{{\bf K}}\,\delta_{{\bf K,K}'}\,\delta_{{\bf r,r}'}.
\end{eqnarray}
Assuming the grand canonical form
\begin{eqnarray} \label{eq2.24}
\rho_s( T,\mu ) = \exp \{ -( H_a-\mu N)/kT\} ,
\end{eqnarray}
which is of quantum Gaussian form, we can easily see that
\begin{eqnarray} \label{eq2.25}
\bar{N}_K( T,\mu ) =\left[ \exp \left(\hbar \omega_{{\bf K}}-
\mu\over kT\right) -1 \right]^{-1}.
\end{eqnarray}
The field operator correlation function is then
\begin{eqnarray} \label{eq2.26}
\langle \psi_{{\bf K}}^{\dagger }( {\bf x}) \psi_{{\bf K}
'}( {\bf x}') \rangle =\delta_{{\bf KK}'}g( {\bf x-x}')
\bar{N}_{{\bf K}}( T,\mu )
\end{eqnarray}
and similarly
\begin{eqnarray} \label{eq2.28}
\langle \psi_{{\bf K}}({\bf x})\psi_{{\bf K}'}^{\dagger }( {\bf x}')\rangle
=\delta_{{\bf K,K}'}g({\bf x-x}')[ 1+N_{{\bf K}}(T,\mu)] .
\end{eqnarray}

\section{Simplest applications of the quantum kinetic master equation}
\label{Sect. 6.5.2}
The situation in which something like the ``molecular chaos'' assumption of
Boltzmann is valid is always of great interest, and it guides the intuition of
most physicists in kinetic theory.  We will consider in this section two
simplified versions of the QKME which are valid in this situation, and one
equation which permits the existence of coherent effects, which cannot arise in
a molecular chaos regime.

The Uehling-Uhlenbeck equation, sometimes called the quantum Boltzmann
equation, is a simple modification of the Boltzmann equation in which the phase
space effects arising from quantum statistics are included.  This equation will
be derived by assuming the distribution is not very different from equilibrium,
and by factorizing higher order correlation functions.

The situation in which we eliminate all dependence on position is obviously an
extreme simplification, and corresponds to the system being composed of only
one spatial cell.  However, in this situation the QKME reduces to a stochastic
master equation of great simplicity and intuitive appeal.  This equation does
not appear ever to have been treated in the literature, so we have given it the
name ``quantum Boltzmann master equation''.

Finally, a very simplified quantum mechanical master equation for the
$ {\bf K}=0$ band can be derived by assuming all other bands are thermalized.
This equation will be called the ``condensate Master equation'', and we are
able to give  an appealing description of the initiation of the condensation
process, and of the growth of the condensate as long as the weak condensation
condition is valid.
\subsection{The quantum Boltzmann master equation}
A very simple stochastic master equation can be derived by assuming that
$ l=\pi/\Delta$ is equal to the size of the whole system, so that
$ \bf r$ takes on only one value. In this case we find that the
$ {\bf K}$-bands are all one-dimensional, so that all the commutator terms,
(\ref{eq2.21a}--\ref{eq2.21d}) are zero.  Since in this case there is really no
difference between $ {\bf n}$ and $ {\bf N}$, we can write
\begin{eqnarray}\label{qbm1}
v_{\bf n}(t) =|{\bf n}\rangle\langle{\bf n}| w_{{\bf n}}(t)
\end{eqnarray}
so that $  w_{{\bf n}}(t) $ is the occupation probability of the state
$ {\bf n} $.  We then derive what we will call the {\em quantum Boltzmann
master equation}
\begin{eqnarray}\label{qbm2}
\dot w_{\bf n} & = & -\frac{\pi}{\hbar^2}\sum_{1234}
\delta(\omega_4+\omega_3-\omega_2-\omega_1)|U_{1234}|^2
\nonumber\\
&&\times\Big\{n_1n_2(n_3+1)(n_4+1)[w_{\bf n}-w_{\bf n+e}]
\nonumber \\
&& + (n_1+1)(n_2+1)n_3n_4[ w_{\bf n}-w_{\bf n-e}]\Big\}
\end{eqnarray}
For clarity of argument we will also write (\ref{qbm1}) in the form
\begin{eqnarray}\label{qbm2a}
\dot w({\bf n}) & = &
\sum_{1234}\Big\{t_{1234}^+({\bf n-e})w({\bf  n-e})
-t_{1234}^+({\bf n})w({\bf n})\Big\}
\nonumber \\
& +&\sum_{1234}\Big\{t_{1234}^-({\bf n+e})w({\bf n+e})
-t_{1234}^-({\bf n})w({\bf n})\Big\}
\end{eqnarray}
where $ t_{1234}^{\pm}({\bf n}) = 0 $ unless
$ \omega_1+\omega_2 = \omega_3+\omega_4$, and if this is satisfied
\begin{eqnarray}\label{qbm3}
t_{1234}^+({\bf n}) & = & \gamma_{1234}(n_1+1)(n_2+1)n_3n_4 \nonumber \\
t_{1234}^-({\bf n}) & = & \gamma_{1234}n_1n_2(n_3+1)(n_4+1)
\end{eqnarray}
and
\begin{eqnarray}\label{qbm4}  
\gamma_{1234} & = & \frac{\pi |U_{1234}|^2}{\hbar^2}.
\end{eqnarray}
\subsubsection{The Boltzmann master equation}
If we drop the terms $ 1+n_i$ in the quantum Boltzmann master equation we get 
an equation called
the Boltzmann Master Equation, which has been previously considered by Van 
Kampen
\cite{vK} and Gardiner \cite{SM} based on stochastic arguments.  The equation
we have given includes no streaming
terms, since we have only one spatial cell, and thus our treatment assumes that
the energy levels are discretely spaced.  This means that our normalization
volume is so small that
the broadening of the levels which arises from collisions is significantly less
than the spacing between levels.

\subsubsection{Stationary solutions}
Apart from the ambiguities caused by conserved quantities, the solutions of
equations like (\ref{qbm2}) are unique.  There is a
consistent detailed balance stationary solution, in which
\begin{eqnarray}\label{qbm7}  
t_{1234}^+({\bf n-e})w_s({\bf n-e}) & = & t_{1234}^-({\bf n})w_s({\bf n})
\end{eqnarray}
and from this we see, using the explicit form (\ref{qbm3}) of $ t^{\pm}$,
that
\begin{eqnarray}\label{qbm8}   	
\frac{w_s({\bf n-e})}{w_s({\bf n})} & = & \frac{t_{1234}^{-}({\bf n})}
{t_{1234}^{+}({\bf n-e})} = 1
\end{eqnarray}
so that for all $ {\bf n}$ able to be connected by a collision $ 1+2
\leftrightarrow3+4$,
\begin{eqnarray}\label{qbm9}   	   	
w_s({\bf n}) = \mbox{constant}.
\end{eqnarray}
Thus $ w_s({\bf n})$ is a function only of conserved quantities, total kinetic
energy, momentum and the total number of particles.   We can thus choose the
grand canonical solution given by
\begin{eqnarray}\label{qbm901}
w_s({\bf n}) \propto \exp\left(-{E-\mu N -{\bf u}\cdot{\bf P}\over kT}\right)
\end{eqnarray}
where $ E$ is the total kinetic energy, $ N$ the total number of particles, and
$ {\bf P}$ the total momentum of the particles, while $ {\bf u}$ is the mean
velocity of the system.  The stationary distribution is in this case
factorizable
\begin{eqnarray}\label{qbm902}
w_s({\bf n}) = \prod_i\left(
e^{-{\hbar\omega_i-\mu-\hbar{\bf K}_i\cdot{\bf u}\over kT}}\right)^{n_i}
\end{eqnarray}

\subsubsection{Factorized equation for the mean occupation numbers}
We can straightforwardly derive the equation for
 $ \langle n_a\rangle = \sum_{\bf n}n_{a}w({\bf n})$
{\begin{eqnarray}\label{4.2.41}
\langle \dot n_a\rangle & = &  4\sum_{234}\gamma_{a234} \Big\{
-\langle n_an_2(n_3+1)(n_4+1)\rangle
\nonumber\\
&&\quad +\langle(n_a+1)(n_2+1)n_3n_4\rangle\Big\}
\end{eqnarray}} %
If we assume that the averages inside this equation can be factorized, which
should be valid if we are not too far from equilibrium, and for
compactness set $ \langle n_i \rangle\to n_i $, we get (without the
streaming terms) the Uehling-Uhlenbeck equation \cite{U-U};
{\begin{eqnarray}\label{4.2.4101}
 \dot n_a & = &  4\sum_{234}\gamma_{a234} \Big\{
- n_an_2(n_3+1)(n_4+1)
\nonumber\\
&&\quad +(n_a+1)(n_2+1)n_3n_4\Big\}
\end{eqnarray}}
The quantum Boltzmann master equation is a very simple equation whose
simulation is entirely feasible, and results of such simulations will be
presented elsewhere.  However, as shown by our derivation, it is expected that
it will not give an accurate description of situations in which there is
significant condensation.  Nevertheless it could well give valuable insights
into the region just below the threshold of condensation, which will be
investigated in QKIII.

\subsection{The Uehling-Uhlenbeck equation}
\label{Sect. 6.4}
We will now show that we can derive from the quantum kinetic master
equation (\ref{eq2.21a}--\ref{eq2.21e}) the more conventional kinetic
equation known as the Uehling-Uhlenbeck equation \cite{U-U}. This equation is
essentially the same as (\ref{4.2.4101}), but with streaming terms added as
well, which in our derivation will arise naturally out of the streaming terms
of the QKME.  For this we consider the one particle distribution function to be
defined as
\begin{eqnarray}\label{eq3.1}
f_{\bf K}({\bf x}) &=&\left(\pi\over\Delta\right)^3
{\rm Tr}\{ \rho \psi_{{\bf K}}^{\dagger}( {\bf x})
\psi_{{\bf K}}( {\bf x}) \} \\
\label{eq3.2}
&=&\left(\pi\over\Delta\right)^3
\sum\limits_{{\bf N}}{\rm Tr}\{ v_{{\bf N}}\psi_{{\bf K}}^{\dagger}( {\bf x})
\psi_{{\bf K}}( {\bf x})\}.
\end{eqnarray}
Notice that the definitions (\ref{eq1.14},\ref{eq1.16}) imply that
\begin{eqnarray}\label{eq3.3}
\psi ({\bf x})=\sum\limits_{{\bf K}}e^{-i{\bf K}\cdot{\bf x}}\psi_{{\bf K}}({
\bf x})
\end{eqnarray}
so that the usual probability density function is, because the $ v_{\bf N}$ are
$ K$-diagonal,
\begin{eqnarray}\label{eq3.4}
\left(\Delta\over\pi\right)^3
{\rm Tr}\{ \rho \psi ^{\dagger }( {\bf x}) \psi ( {\bf x}) \} =
\sum\limits_{{
\bf K}} f_{\bf K}({\bf x})
\end{eqnarray}
so that $ f_ {\bf K} ({\bf x}) $ can
be viewed as a phase space density per phase space volume $ h^3$.

\subsubsection{Streaming terms}
We consider the various terms in the quantum kinetic master equation.
The term arising from (\ref{eq2.21a}) is obtained by considering
\begin{eqnarray}\label{eq3.6}
\left[ \psi_{{\bf K}}( {\bf x}) ,\int d^3{\bf y}\,j_{{\bf K}}( {\bf y}) \right]
\end{eqnarray}
which can be written
\begin{eqnarray}\label{eq3.7}
\left[
\psi_{{\bf K}}( {\bf x}),\int d^3{\bf y}\,\psi_{{\bf K}}^{\dagger}({\bf y})
\left( -{i\hbar\over m } \right) \nabla \psi_{{\bf K}}( {\bf y})
\right] \\
\label{eq3.8}
=\int d^3{\bf y}\,g( {\bf x}-{\bf y}) \left( -{i\hbar\over m}\right)
\nabla_{\bf y}\psi_{\bf K}({\bf y})
\end{eqnarray}
and now integrating by parts,
\begin{eqnarray}\label{eq3.9}
(\ref{eq3.8})= -{i\hbar \over m} \nabla_{\bf x}\left(\int d^3{\bf
y\,}g( {\bf x-y}) \psi_{{\bf K}}( {\bf y}) \right)
\end{eqnarray}
but the integral is simply the projection of $ \psi_{{\bf K}}( {\bf
y}) $ on to the $ {\bf K} $ subspace; thus
\begin{eqnarray}\label{eq3.10}
(\ref{eq3.8})= -{i\over\hbar m }\nabla \psi_{{\bf K}}( {\bf x}) .
\end{eqnarray}

Using this kind of technique, it is straightforward to show that the
contributions to $\dot f_{\bf K}({\bf x}) $ from
(\ref{eq2.21a}, \ref{eq2.21e}) become
\begin{eqnarray}\label{eq3.11}
\dot f_{\bf K}({\bf x})|_a &=&
{\hbar {\bf K}\over m}\cdot \nabla_{\bf x} f({\bf K,x})   \\
\label{eq3.12}
\dot f_{\bf K}({\bf x})|_b &=&\left(\pi\over\Delta\right)^3
\nabla \cdot {\bf j}_{{\bf K}}({\bf x})
\end{eqnarray}
where
\begin{eqnarray}\label{eq3.13}
{\bf j}_{{\bf K}}( {\bf x})
= {i\hbar\over 2m}\left\langle\psi_{{\bf K}}^{\dagger }({\bf x})
 \nabla \psi_{{\bf K}}({\bf x}) -[\nabla\psi_{{\bf K}}^{\dagger}({\bf x})]
\psi_{{\bf K}}({\bf x})\right\rangle  .
\end{eqnarray}
The term (\ref{eq3.11}) is the usual Boltzmann streaming
term. The term (\ref{eq3.12}) is a purely quantum mechanical term
arising from the spreading of the wavepacket.

In order to get some idea of the significance of this and some terms
to come, we make a {\em local equilibrium ansatz}, in which we write
\begin{eqnarray}\label{eq3.14}
\langle \psi_{{\bf K}}^{\dagger }( {\bf x}) \psi_{{\bf K}'}( {\bf x}')
\rangle =g( {\bf x-x}') \;f_{\bf K}\left(\frac{{\bf x+x}'}2\right)
\delta_{{\bf K,K}'}\,.
\end{eqnarray}
This is a natural generalization of (\ref{eq2.26}) to a slightly non
equilibrium situation, and since $ g(0)=(\pi/\Delta)^3$, it is consistent with
(\ref{eq3.1}).  Using this form we find quite
straightforwardly that $ {\bf j}_{{\bf K}}( {\bf x}) =0. $ The extent
to which this term does not vanish must be related to the extent to
which the mean momentum in the band
$ ( {\bf K}-{\bf \Delta},\,{\bf K}+{\bf \Delta} ) $ is different from
$ \hbar {\bf K} $, and this difference
is of order of magnitude $ \Delta $, which will only be appreciable
for very small $ {\bf K} $, and in particular for $ {\bf K}=0$.

\subsubsection{Forward scattering terms}
If we work out the mean value of the commutator, we find for the term
arising from  (\ref{eq2.21c})
\begin{eqnarray}\label{eq3.15}
&&-{ i\over\hbar} \Big\{ 2\int d^3{\bf x}'\sum\limits_{{\bf K}'}
\langle \psi_{{\bf K}}^{\dagger }( {\bf x}') \psi_{{\bf K}'}^{\dagger }( {\bf
x}')\psi_{{\bf K}'}( {\bf x}') \psi_{{\bf K}}( {\bf x}) \rangle g(
{\bf x}-{
\bf x}') . \nonumber \\
&&\qquad -2\int d^3{\bf x}'\langle \psi_{{\bf K}}^{\dagger}( {\bf x})
\psi_{{\bf K}'}^{\dagger }( {\bf x}') \psi_{{\bf K}'}( {\bf
x}')\psi_{{\bf K}}( {\bf x}') \rangle g( {\bf x-x}') \Big\} .
\nonumber \\
\end{eqnarray}
If $ g( {\bf x-x}') $ was a perfect delta function, this term would
vanish, thus the degree to which it does not vanish depends on the
smoothness of the averages as functions of $ {\bf x,x}' $.

Using the local equilibrium ansatz (\ref{eq3.14}), and a Gaussian
factorization assumption on the four point correlations, it is easy to
show that this term vanishes, and is therefore presumably very small
when the state is close to equilibrium.  Thus we can say
$ \dot f_{\bf K}({\bf x})|_c\approx 0  $.
\subsubsection{Collision terms}
In evaluating the contribution from  (\ref{eq2.21e}) we
will get a number of terms, of which a typical one is proportional to
\begin{eqnarray} \label{eq3.16}
&&-\sum\limits_{{\bf K}_2,{\bf K}_3,{\bf K}_4}\int d^3{\bf y}\,d^3{\bf y}'
\,g( {\bf x-y}')
\langle\psi_{{\bf K}}^{\dagger}( {\bf x})\psi_{{\bf K}_2}^{\dagger}( {\bf y}')
\psi_{{\bf K}_3}({\bf y}') \psi_{{\bf K}_4}( {\bf y}')
 \psi_{{\bf K}_4}^{\dagger }( {\bf y}) \psi_{{\bf K}_3}^{\dagger }(
{\bf y}) \psi_{{\bf K}_2}( {\bf y}) \psi_{{\bf K}}^{\dagger }( {\bf
x}) \rangle
\nonumber \\
 &&\qquad\times \delta (\omega_{{\bf
K}}+\omega_{{\bf K}_2}-\omega_{{\bf K}_3} -\omega_{{\bf K}_4}) e^{i(
{\bf K}+{\bf K}_2-{\bf K}_3-{\bf K}_4) \cdot ( {\bf y-y}') }.
\end{eqnarray}
If we factorize as in local equilibrium, and use (\ref{eq3.14}), we
get terms like
\begin{eqnarray} \label{eq3.17}
&&\sum\limits_{{\bf K}_2,{\bf K}_3,{\bf K}_4}\int d^3{\bf y}\,d^3{\bf
y}'\, g( {\bf x-y}') g( {\bf x-y})[ g( {\bf y}'-{\bf y}) ] ^3 f_{{\bf
K}}\left(
\frac{{
\bf x}+{\bf y}}2\right)f_{{\bf K}_2}\left(\frac{{\bf y}'+{\bf y}}2\right)
\nonumber \\
&&\quad\times [f_{{\bf K}_3}\left(\frac{{\bf y}'+{\bf y}}2\right)+1]
[f_{{\bf K}_4}\left(\frac{{\bf y}'+{\bf y}}2\right) +1]
\delta (\omega_{{\bf K}}+\omega_{{\bf K}_2}-\omega_{{\bf
K} _3}-\omega_{{\bf K}_4}) e^{i( {\bf K}+{\bf K}_2-{\bf K}_3-{\bf
K}_4) \cdot ( {\bf y-y}') }.
\end{eqnarray}
Assuming the functions are relatively smooth, so that we can set
(because of the peaked nature of $ g( {\bf x-y}')$, $ g( {\bf x-y}) $ and
$ g( {\bf y}'-{\bf y}) $
\begin{eqnarray} \label{eq3.18}
{{\bf x}+{\bf y}\over 2}\approx{{\bf y}'{\bf +y}\over 2}\approx{\bf x}
\end{eqnarray}
we get
\begin{eqnarray} \label{eq3.19}
&&\sum\limits_{{\bf K}_2,{\bf K}_3,{\bf K}_4} f_{{\bf K}}( {\bf x})
f_{{\bf K}_2}( {\bf x}) [f_{{\bf K}_3}( {\bf x})+1][ f_{{\bf K}_4}(
{\bf x})+1]
\delta (\omega_{{\bf K}}+\omega_{{\bf K}_2}-\omega_{{\bf K}_3}-\omega_{{\bf
K}_4}) \nonumber\\ &&\qquad\times \int d^3{\bf y}\int d^3{\bf y}' g(
{\bf y}') g( {\bf y}) [ g( {
\bf y}'-{\bf y}) ]^3 e^{i( {\bf K}+{\bf K}_2-{\bf K}_3-{\bf K}_4)\cdot ( {\bf
y-y}') }.
\end{eqnarray}
The last factor is in fact able to be evaluated exactly, and the result is
a momentum conservation function
$ M_\Delta({\bf K+K}_2-{\bf K}_3-{\bf K}_4)$, where
\begin{eqnarray} \label{eq3.20}
M_{\Delta \,}( {\bf Q}) =\left({\Delta}\over{\pi  }\right)^3
\prod\limits_{i=1}^3 \{ {\mbox{$ 2\over3$}}\delta_{Q_i,0}
+ {\mbox{$ 1\over6$}}\delta_{Q_i,\Delta }+{\mbox{$
1\over6$}}\delta_{Q_i,-\Delta }\}.
\end{eqnarray}
Thus ``momentum'' $ {\bf K} $ is not exactly conserved, since the
quantity ${\bf K} $ in{\bf } $ \psi_{{\bf K}}( {\bf x}) $ represents
only the midpoint of the range of possible momenta; it is possible for
three momenta, each from such a range, to add up to a value either
within the range, or in either of the ranges above or below. The exact
proportions of each are shown by the calculation to be $ {2\over
3}:{1\over 6}:{1\over 6} $.  However, assuming that $ f_{{\bf K}}(
{\bf x}) $ are smooth functions of $ {\bf K} $ on the scale of $
\Delta $, we can ignore the slight spreading, and make the transfer
from a summation to an integration: This yields
\begin{eqnarray}\label{eq3.21}
\sum\limits_{{\bf K}_2,{\bf K}_3,{\bf K}_4}
\left({\Delta}\over{\pi  }\right)^3
M_\Delta ( {\bf K+K}_2-{\bf K}_3-{\bf K}_4)
\to {1\over( 2\pi ) ^3}\int d^3{\bf K}_2\int \,d^3{\bf K}_3
\int \,d^3{\bf K}_4\int \delta ( {\bf K}+{\bf K}_2-{\bf K}_3-{\bf K}_4)
\nonumber \\
\end{eqnarray}
so that this form eventually becomes
\begin{eqnarray}\label{eq3.22}
-{u\over 2\hbar^2}\int d^3{\bf K}_2\,\int d^3{\bf K}_3\int \,d^3{\bf
K}_4\
\delta ( {\bf K}+{\bf K}_2-{\bf K}_3-{\bf K}_4)
 \delta (\omega +\omega_2-\omega_3-\omega_4) &&
  f_{{\bf K}}( {\bf x}) f_{{\bf K}_2}( {\bf x})
[ f_{{\bf K}_3}( {\bf x})+1][ f_{{\bf K}_4}( {\bf x})+1]
\nonumber\\
\end{eqnarray}
Here we have made the approximation
\begin{eqnarray}\label{eq3.23}
u({\bf x-x}')=  u  \delta ( {\bf x-x'})
\end{eqnarray}
Taking all terms into account we eventually get a term
\begin{eqnarray}\label{eq3.24}
\dot f( {\bf K,x})|_e  & = &
\frac{2| u| ^2}{h^2}\int d^3{\bf K}_2\int d^3{\bf
K}_3\int d^3{\bf K}_4 \delta ( {\bf K}+{\bf K}_2-{\bf K}_3-{\bf K}_4)
\delta (\omega +\omega_2-\omega_3-\omega_4)
\nonumber \\
&&\times\{f_{{\bf K}}( {\bf x}) f_{{\bf K}_2}( {\bf x})
[ f_{{\bf K}_3}({\bf x}) +1] [ f_{{\bf K}_4}( {\bf x}) +1]
- [ f_{{\bf K}}( {\bf x}) +1] [ f_{{\bf K}_2}( {\bf x})+1]
f_{{\bf K}_3}({\bf x})f_{{\bf K}_4}({\bf x}) \}
\end{eqnarray}
which is the collision term of the equation known as the
Uehling-Uhlenbeck equation.

\subsubsection{Summary}

In the situation that we have an approximate local equilibrium, and
that multifield correlation functions can be factorized as if the
density operator is Gaussian, we derive the Uehling-Uhlenbeck
equation as an approximate kinetic equation.
\begin{eqnarray}\label{eq3.25}
{\partial f\over \partial t}( {\bf K,x}) &\approx&
 \frac{\hbar {\bf K}\cdot \nabla_{{\bf x}}}m f( {\bf K,x})
 +\frac{2| u|^2}{h^2}\int d^3{\bf K}_2\,\int d^3{\bf K}_3\int
\,d^3{\bf K}_4 \delta ( {\bf K}+{\bf K}_2-{\bf K}_3-{\bf K}_4)
\delta (\omega +\omega_2-\omega_3-\omega_4)
\nonumber \\
&&\times \{
f_{{\bf K}}( {\bf x}) f_{{\bf K}_2}( {\bf x} )
[ f_{{\bf K}_3}( {\bf x}) +1] [ f_{ {\bf K}_4}( {\bf x}) +1]
 - [f_{{\bf K}}( {\bf x}) +1] [ f_{{\bf K}_2}( {\bf x}) +1]
 f_{{\bf K}_3}({\bf x}) f_{{\bf K}_4}( {\bf x}) \}
\nonumber\\
\end{eqnarray}
In situations of interest for Bose Condensation it would no longer be
valid to make the factorizations necessary for the collision term as
above, nor factorization in the local equilibrium assumption
(\ref{eq3.14}).

\subsection{\label{Condeq}Equations for the condensate density operator}
Let us now assume that the scattering of non condensate modes is strong,
and that we are only interested in the behavior of the condensate. We can
then assume that the total density operator can be factored into a
thermal non-condensate part, and a condensate density operator
$ \rho_0(t) $ as a first approximation at least. Thus
\begin{eqnarray}\label{eq4.10}
\rho ( t) =\sum\limits_{{\bf N}}v_{{\bf N}}(t) \to
 \rho _B(\mu, T) \otimes \rho_0( t)
\end{eqnarray}
This situation is not that employed in current experiments, where the
condensate grows by taking atoms out of the bath of warmer atoms, but is given
for illustrative purposes only---it describes the situation in which the
condensate mode is put in contact with a bath composed of all the atoms in the
other modes, which are held at a given temperature and chemical potential.
More realistic treatments of condensate growth are left to a later paper.

We trace over the non-condensate modes and the assumption that the 
non-condensate is thermalized will mean that all terms not involving condensate  
operators in the QKME will be vanish.  The only non-vanishing dissipative terms 
will involve matched $\psi _0^{\dagger }$ and $ \psi _0 $ operators, so that 
we 
arrive at the {\em condensate master equation}:
\begin{eqnarray}\label{eq4.11}
\dot \rho _0( t)  &=&-{i\over\hbar}  \left[ \int d^3{\bf x}\,
\psi_0^{\dagger }( {\bf x}) \left( -{\hbar ^2\nabla ^2\over 2m}\right)
\psi _0({\bf x}) +{1\over 2}U( 0000) ,\rho_0( t) \right]
-{i\over\hbar}(u g(0) \sum\limits_{{\bf K}\neq 0}\bar n_{{\bf K}})
\left [ \int d^3{\bf x}\,\psi _0^{\dagger }( {\bf x}) \psi _0( {\bf x}) ,\rho_0
\right]
\nonumber \\
&&+\int d^3{\bf x}\int d^3{\bf x}'G^{(-) }( {\bf x-x}',T,\mu)
 \{ 2\psi _0( {\bf x}) \rho_0\psi _0^{\dagger }( {\bf x}') -\rho
_0\psi _0^{\dagger}( {\bf x}') \psi _0( {\bf x}) -\psi_0^{\dagger }(
{\bf x}') \psi _0( {\bf x})\rho _0\}
 \nonumber \\
 &&+\int d^3{\bf
x}\int d^3{\bf x}'G^{( +) }( {\bf x-x}',T,\mu)
\{ 2\psi_0^{\dagger }( {\bf x}) \rho _0\psi _0( {\bf x}') -\rho _0\psi _0({\bf
x}') \psi _0^{\dagger }( {\bf x}) -\psi_0( {\bf x}') \psi _0^{\dagger
}( {\bf x})\rho _0\} , 
\end{eqnarray}
in which the quantities $ G^{(\pm ) } $ are given by
\begin{eqnarray}\label{eq4.12}
G^{( +) }( {\bf x-x}',T,\mu) &=&{\rm Tr}_B\left\{ \sum\limits_{{123}}
{\pi\over\hbar^2}Z^{\dagger}( 1,2,3,{\bf x})Z(1,2,3,{\bf x}') \rho _B(T,\mu)
\right\}
\nonumber \\
G^{( -) }( {\bf x-x}',T,\mu) &=&{\rm Tr}_B\left\{ \sum\limits_{{123}}
{\pi\over\hbar^2}Z( 1,2,3,{\bf x}) Z^{\dagger }( 1,2,3,{\bf x}') \rho _B(T,\mu)
\right\} .
\nonumber \\
\end{eqnarray}
with the definition
\begin{eqnarray}\label{eq4.1301}
Z({\bf K}_1,{\bf K}_2,{\bf K}_3,{\bf x})
= u e^{-i({\bf K}_1+{\bf K}_2-{\bf K}_3)\cdot {\bf x}}
\psi_{{\bf K}_1}({\bf x})
\psi_{{\bf K}_2}( {\bf x})
\psi_{{\bf K}_3}^{\dagger }({\bf x})
\end{eqnarray}
Using thermal averages we find
\begin{eqnarray}\label{eq4.13}
G^{( +) }( {\bf x-x}',T,\mu) &=&
\frac{\pi u^2}{\hbar ^2}\{\sum\limits_{{123}}
[ g( {\bf x-x}') ]^3 e^{i( {\bf K}_1{\bf +K}_2{\bf-K}_3)\cdot ( {\bf x-x}')}
\bar{n}_{{\bf K}_1}\bar{n}_{{\bf K}_2}(\bar{n}_{{\bf K}_3}+1) \delta
(\omega_1+\omega _2-\omega _3) \}
\nonumber \\ G^{( -) }( {\bf x-x}',T,\mu) &=&
\frac{\pi u^2}{\hbar ^2}\{
\sum\limits_{{123}}[ g( {\bf x-x}') ] ^3 e^{i( {\bf K}_1{\bf +K}_2{\bf
-K}_3)\cdot ( {\bf x-x}') }
(\bar{n}_{{\bf K}_1}+1)(\bar{n}_{{\bf K}_2}+1) \bar{n}_{{\bf K}_3}
\delta (\omega _1+\omega _2-\omega _3) \}
\end{eqnarray}
Suppose we now set, as in (\ref{eq2.25})
\begin{eqnarray}\label{eq4.14}
\bar{n}_{\bf K}=\left( e^{(\hbar \omega _{\bf K} -\mu )/kT} - 1 \right )^{-1},
\qquad( {\bf K\neq }0)
\end{eqnarray}
then it follows that
\begin{eqnarray}\label{eq4.15}
\bar{n}_{{\bf K}_1}\bar{n}_{{\bf K}_2}( 1+\bar{n}_{{\bf K}_3})
=e^{\mu /kT}(\bar{n}_{{\bf K}_1}+1) (\bar{n}_{{\bf K}_2}+1)
\bar{n}_{{\bf K}_3},
\end{eqnarray}
so that
\begin{eqnarray}\label{eq4.16}
G^{(+) }( {\bf x-x}',T,\mu) =e^{\mu/kT}G^{(-) }( {\bf x-x}',T,\mu) .
\end{eqnarray}

\subsubsection{Quantum stochastic differential equation form}
We can write a Quantum stochastic differential equation (QSDE) equivalent to
the master equation (\ref{eq4.11}) by using the methods in Chap.5 of
\cite{Gardiner}.  This equation takes the form
\begin{eqnarray}\label{qs1}
i\hbar\, d\psi_0({\bf y},t) &=&
\Bigg\{ -{\hbar^2\over 2m}\nabla^2\psi_0({\bf y},t)
-\left(
u g(0)\sum\limits_{{\bf K}\neq0}\bar{n}_{{\bf K}}\right)\psi_0({\bf y},t)
+u\int d^3{\bf x}\,g({\bf x}-{\bf y})
\psi_0^{\dagger}({\bf x},t)\psi_0( {\bf x},t)\psi_0({\bf x},t)
\nonumber \\
&&+\int d^3{\bf x}\,\int d^3{\bf x}'g( {\bf x}'-{\bf y})
\{ G^{(+) }( {\bf x-x}',T,\mu )
-G^{(-) }( {\bf x-x}',T,\mu ) \}\psi_0( {\bf x},t) \Bigg\}
\nonumber \\
&&+ \int d^3{\bf x}\, g({\bf x}-{\bf y})\,dW({\bf x},t)
\end{eqnarray}
in which $ dW({\bf x},t) $ is a quantum white noise increment which satisfies
the conditions
\begin{mathletters}
\begin{eqnarray}\label{qs2a}
dW({\bf x},t)dW^\dagger({\bf x}',t) &=& 2 G^{(-) }( {\bf x-x}',T,\mu )
\\ \label{qs2b}
dW^\dagger({\bf x},t)dW({\bf x}',t) &=& 2 G^{(+) }( {\bf x-x}',T,\mu )
\\ \label{qs2c}
dW({\bf x},t)dW({\bf x}',t)&=&0
\\ \label{qs2d}
dW^\dagger({\bf x},t)dW^\dagger({\bf x}',t)&=&0
\end{eqnarray}
\end{mathletters}

\subsubsection{Gain and loss}\label{Sect. 6.5.3}
In order to get some idea of the predictions of the master equation
(\ref {eq4.11}) we can write equations for the averages
\begin{eqnarray}\label{eq4.17}
\langle \psi _0( {\bf y}) \rangle &\equiv& \phi ({\bf y}) ,  \\
 \label{eq4.18}
\langle \psi _0^{\dagger }( {\bf y}) \psi _0( {\bf y}) \rangle
& \equiv& \bar \rho ( {\bf y}) .
\end{eqnarray}
These become
\begin{eqnarray} \label{eq4.19}
\frac{{\cal \partial }\phi ( {\bf y}) }{\partial t} &= &
-\frac i\hbar \left\{ -\frac{\hbar ^2}{2m}\nabla ^2\phi ( {\bf y})
-\left(u g(0)\sum\limits_{{\bf K}\neq 0}\bar{n}_{{\bf K}}\right)\phi ({\bf y})
+u\int d^3{\bf x}\,g({\bf x}-{\bf y})
\langle\psi_0^{\dagger}({\bf x})\psi_0( {\bf x})\psi_0({\bf x})\rangle \right
\}
\nonumber \\
&&+\int d^3{\bf x}\,\int d^3{\bf x}'g( {\bf x}'-{\bf y})
\{ G^{( +) }( {\bf x-x}',T,\mu )
-G^{( -) }( {\bf x-x}',T,\mu ) \} \phi( {\bf x})
\nonumber \\
\label{eq4.20}
\frac{{\cal \partial }\bar{\rho }( {\bf y}) }{\partial t}
&=&  -\nabla \cdot {\bf j}( {\bf y}) +
\int d^3{\bf x}\{G^{( +) }( {\bf y-x},T,\mu ) - G^{( -)}( {\bf y-x},T,\mu ) \}
\nonumber \\
&&\times \langle \psi _0^{\dagger }( {\bf x})
 \psi _0({\bf y}) +\psi _0^{\dagger }( {\bf y}) \psi _0( {\bf x})\rangle
+2\int d^3{\bf x\ }G^{( +) }( {\bf x},T,\mu ) g( {\bf x}).
\end{eqnarray}
Here $ {\bf j}( {\bf y}) $ is the probability current. The equation
(\ref{eq4.19}) for $ \phi ( {\bf y}) $ is a rather familiar non-linear
Schr\"odinger equation form, but contains the second line as well,
which is an explicit gain-loss term. We can use the relation
(\ref{eq4.16}) to write this in the form
\begin{eqnarray}\label{eq4.21}
\int d^3{\bf x}\int d^3{\bf x}'g( {\bf x}'{\bf -y})
( e^{\mu /kT}-1) G^{( +) }( {\bf x-x}',T,\mu ) \phi ( {\bf y})
\end{eqnarray}
and clearly when $ \mu =0 $, this vanishes.

If we consider spatially homogeneous solutions, then we can write 
$ \phi ({\bf x})\rightarrow\phi$, and we find
\begin{eqnarray}\label{eq4.22}
\frac{\partial \phi}{\partial t}& =&
 -\frac{i}{\hbar}\bigg(ug(0)\sum _{{\bf k}\not= 0}
\bar{n}_{\bf K}\bigg)\phi-\frac{i}{\hbar}u
\int d^3{\bf x}\,g({\bf x}-{\bf y})
\langle \psi^{\dagger}_{0}({\bf x})\psi_{0}({\bf x})\psi_{0}({\bf x})\rangle
\nonumber \\
&& + \phi(e^{\mu/kT}-1) \frac{u^2}{2\hbar^{2}\pi^{5}}\int d^{3}{\bf
K}_{1}\int d^{3} {\bf K}_{2}\bar {n}_{{\bf K}_1}\bar {n}_{{\bf K}_2}
 (1+\bar {n}_{{\bf K}_1+{\bf K}_2})\delta(\omega_{{\bf K}_{1}}+\omega_{{\bf
K}_{2}}-\omega_{{\bf K}_{1}+{\bf  K}_{2}}).
\nonumber \\
\end{eqnarray}
\subsubsection{Stationary solutions.}
If $ \mu <0$, as must be the case above the condensation
temperature, the last part clearly represents a loss term.  There will be no
growth of a non zero $ \phi$ value.  If we assume a Gaussian solution, we
can write
\begin{eqnarray}\label{eq4.23}
\langle\psi^{\dagger}_{0}({\bf y})\psi_{0}({\bf y})\psi_{0}({\bf y})\rangle
\rightarrow 2 \phi\langle
\psi^{\dagger}_{0}({\bf y})\psi_{0}({\bf y})\rangle
\end{eqnarray}
and it is clear that the only stationary solution will be $\phi=0$.

Similarly, we can get an approximate closed equation from (\ref{eq4.20}) by
assuming
\begin{eqnarray}\label{eq4.24}
\langle \psi^{\dagger}_{0}({\bf x})\psi_{0}({\bf y})\rangle = \frac{g({\bf x}-{
\bf
y})}{g(0)}\bar \rho ,
\end{eqnarray}
which assumes the equilibrium {\it shape} (\ref{eq2.28}) of the correlation
function.  This leads to
\begin{eqnarray}\label{eq4.25}
\frac {\partial\bar {\rho}}{\partial t} 
= 2 \int d^{3}{\bf x}G^{+}({\bf y}-{\bf x},T,\mu)g({\bf x}-{\bf y})
\big\{ (e^{\mu/kT}- 1)\bar {\rho}/g(0) + 1 \big\} .
\end{eqnarray}
We then find the stationary solution
\begin{eqnarray}\label{eq4.26}
\bar {\rho} = \frac{g(0)}{1-e^{\mu/kT}} = \frac{l^{-3}}{1-e^{\mu/kT}},
\end{eqnarray}
which gives the total mean number per volume as $ l^{3}$ as $1/(1-e^{\mu/kT})
$, the usual statistical mechanical result when $ \mu <0$.  When $ \mu=0$
there
is no longer a stationary state of Eq..(\ref{eq4.20}); ${\bar \rho} $ will grow
indefinitely.  However there is still no gain, so no coherent phase appears.
Of course in practice $ \mu$ is {\it not} exactly zero, so indefinite growth
will not occur.

Thus we see that the ideal Bose condensate result arises purely out of
{\it noise}; that no coherent phase appears at all.  This result is only valid
in the weak condensation limit; we will see in a subsequent paper that higher
order terms can lead to an effective gain, and that a non-zero stationary value
of $ \phi$ can arise.

\subsubsection{Gain and a coherent phase arising from a modified
distribution of energies in the non-condensate}
It is of interest to see whether we could produce a non-zero value of $ \phi$
by manipulating the nature of the energy distribution of the non-condensate.
If we do  assume that the non-condensing
modes are in a non-equilibrium stationary state, so that instead of $ \rho_B(T,
\mu)$ we have an arbitrary density operator $ \rho_{\rm ne}$,
the gain in (\ref{eq4.19}) becomes
\widetext\onecolumn
\begin{eqnarray}\label{eq4.27}
{u^{2}\over2\hbar^{2}\pi^{5}}\int d^{3}{\bf K}_{1}d^{3}{\bf K}_{2}
\delta
(\omega_{{\bf K}_{1}}+\omega_{{\bf K}_{2}}-\omega_{{\bf K}_{1}+{\bf K}_{2}})
\bigg\{
\bar{n}_{{\bf K}_1}\bar{n}_{{\bf K}_2}[1+\bar{n}_{{\bf K}_1+{\bf K}_2}]
-[1+\bar{n}_{{\bf K}_1}][1+\bar{n}_{{\bf K}_2}] \bar{n}_{{\bf K}_1+{\bf K}_2}
\bigg\}.
\nonumber\\
\end{eqnarray}
We can get gain if (but not only if) every team in the $ \{ \}$ is positive.
Assuming that $\bar {n}({\bf K})$ is only a function of $\omega_{{\bf K}} $,
this
will happen when
\begin{eqnarray}\label{eq4.28}
\frac{\bar{n}(\omega_{1})}{1+\bar{n}(\omega_{1})}\,
\frac{\bar{n}(\omega_{2})}{1+\bar{n}(\omega_{2})}\,
 \frac{1+\bar{n}(\omega_{1}+\omega_{2})}{\bar{n}(\omega_{1}+\omega_{2})} > 1
\end{eqnarray}
or, defining  \begin{eqnarray}\label{eq4.29}
F(\omega) = \log\left\{\frac{\bar{n}(\omega)}{1+\bar{n}(\omega)}\right\}
\end{eqnarray}
\begin{eqnarray}\label{eq4.30}
F(\omega_{1}) + F(\omega_{2}) > F(\omega_{1}+\omega_{2}).
\end{eqnarray}
This is the condition that $F(\omega) $ is a convex downwards function
as in the solid line in the diagram.  For example,
\begin{eqnarray}\label{eq4.31}
 F(\omega) = \frac {-\hbar\omega- \lambda\omega^{2}}{kT}
\end{eqnarray}
gives \begin{eqnarray}
 F(\omega_{1})+F(\omega_{2})-F(\omega_{1}+\omega_{2})=\frac{2\lambda\omega_{1}
\omega_{2}}{kT}>0
\end{eqnarray}
and the net gain is
\begin{eqnarray}\label{eq4.32}
\frac{u^{2}}{2\hbar^{2}\pi^{5}}\int d^{3}{\bf K}_{1}d^{3}{\bf K}_{2}\delta(
\omega_{\bf K_{1}}+\omega_{\bf K_{2}}-\omega_{{\bf K}_{1}+{\bf K}_{2}})
\nonumber \\
\times\exp\left(\frac{2\lambda\omega_{\bf K_{1}}\omega_{\bf K_{2}}}{kT}\right)
\bar{n}_{\bf K_{1}}\bar{n}_{\bf K_{2}}[1+\bar{n}_{{\bf K}_1+{\bf K}_2}].
\end{eqnarray}

\section{Conclusions}
Quantum kinetic theory is both a genuine kinetic theory
and a genuine quantum theory.  The kinetic aspect arises from the decorrelation
between different momentum bands, and this is essentially our version of the
molecular chaos condition.  However, because each
${\bf K}$-band has a range
of momenta within it, from which coherences are not eliminated, there is still
the possibility of long wavelength quantum coherence.  In practice such
coherence is not expected to be of significance except in the Bose condensate,
but the formulation we have developed enables a correct formulation of the
method by which incoherence can be transferred into the condensate.

We have formulated quantum kinetic theory in terms of the Quantum Kinetic
Master Equation for bosonic atoms. It is a quantum stochastic equation for the
kinetics of a dilute quantum degenerate bose gas, and it is valid in the regime
of Bose condensation, as well as in regimes where no condensation takes place.

The QKME is a genuine $N$-atom equation, which is intermediate between the full
description in terms of the complete density matrix and the familiar kinetic
equations for single particle distribution functions (like the quantum
Boltzmann equation). It contains as limiting cases both the Uehling-Uhlenbeck
(or quantum Boltzmann equation) and the Gross-Pitaevskii equation.

The key assumption in deriving the QKME is a Markov approximation for the
atomic collision terms. This is valid to the extent we are interested in a
situation where in addition to the Bose condensate we have a thermal fraction
of atoms which effectively acts like a heat bath with a (short) thermal
correlation time $\hbar/kT$, which is taken as zero in the Markov
approximation.

The present paper has developed the basic structure of the theory,
has stated and justified the approximations which are needed, and has thus
delineated the region of validity of the theory.
Extensions to include a trapping potential and to account for a large fraction
of condensed atoms giving rise to Bogoliubov type excitation spectrum are
essentially technical, because the basic formulation carries through with
little conceptual change, although the equations which are appropriate in these
cases are significantly changed.
These aspects, which are essential for the application of the QKME method to
realistic experiments, will be dealt with in QKII\cite{QKII}.

Acknowledgment: We thank K. Burnett and J.I. Cirac for discussions. C.W.G.  
thanks the Marsden Fund for financial support under contract number GDN-501. 
P.Z. was supported in part by the \"Osterreichische Fonds zur F\"orderung der 
wissenschaftlichen Forschung.


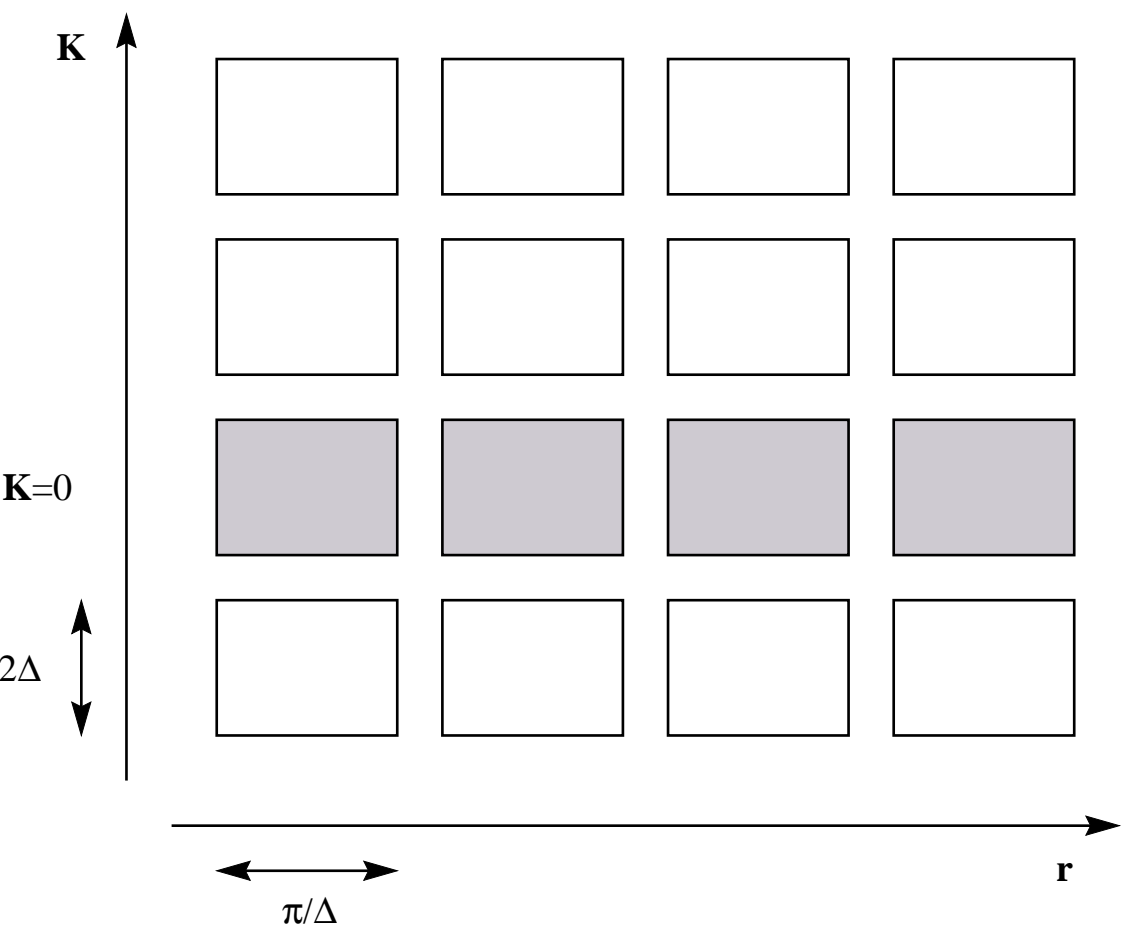
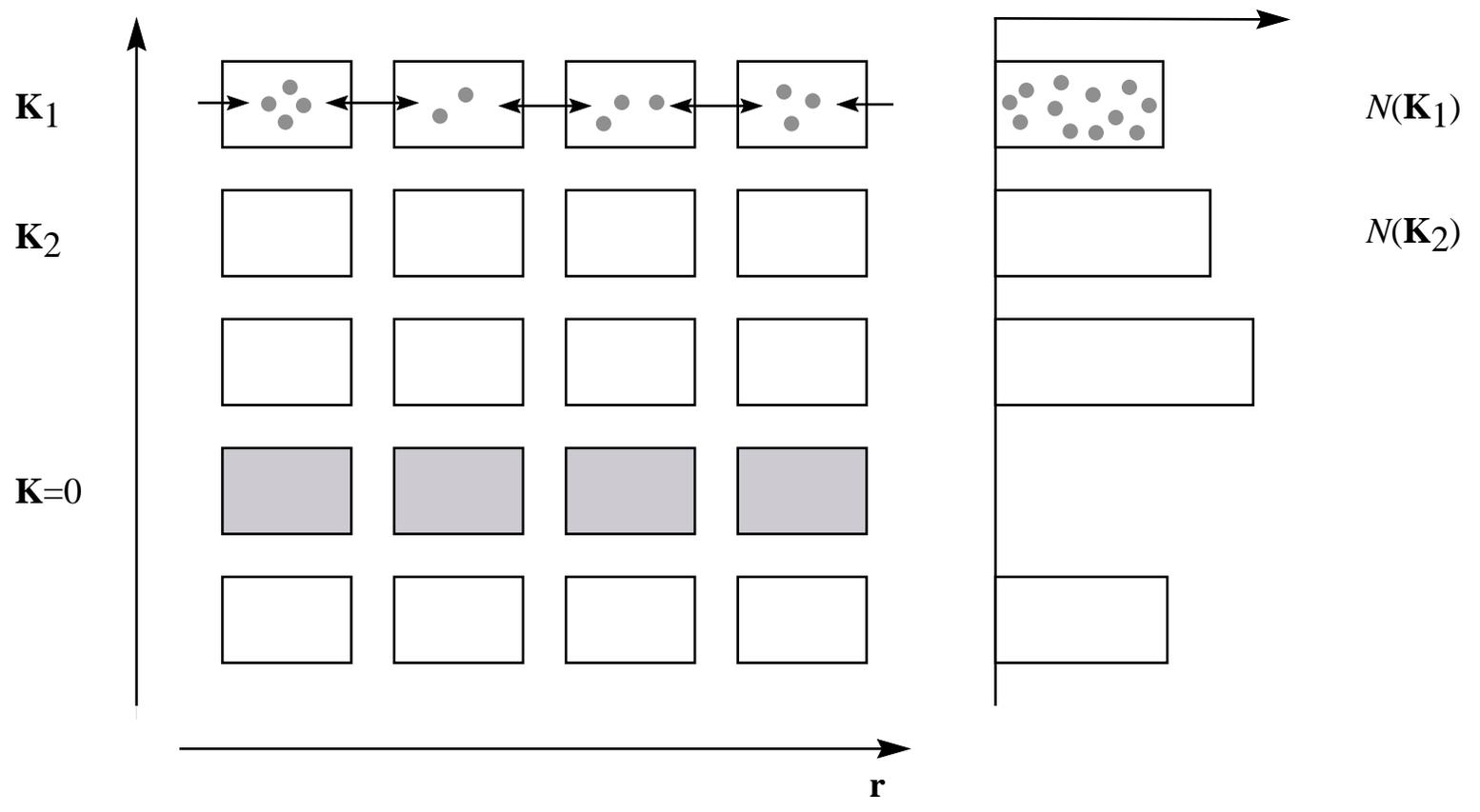
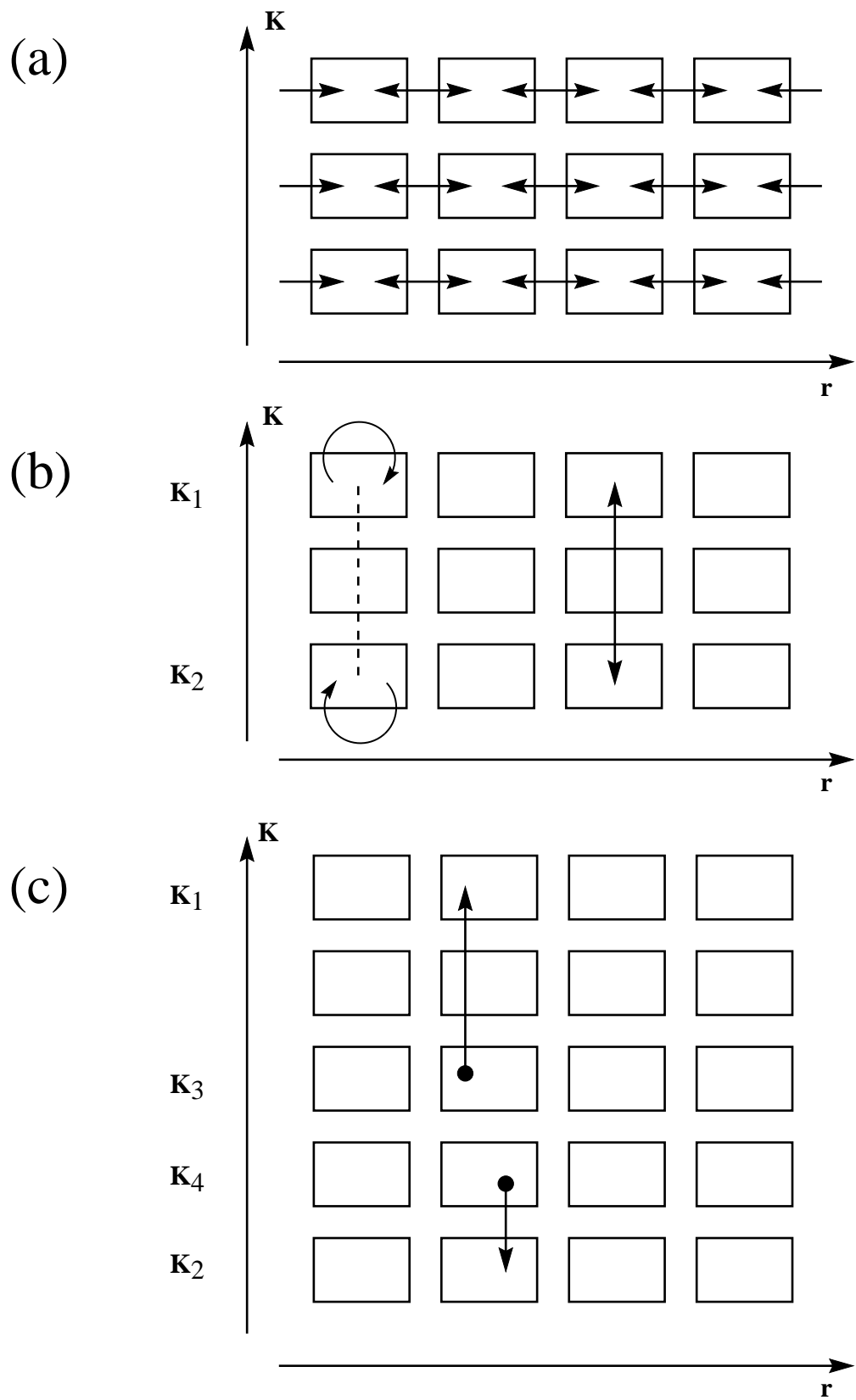
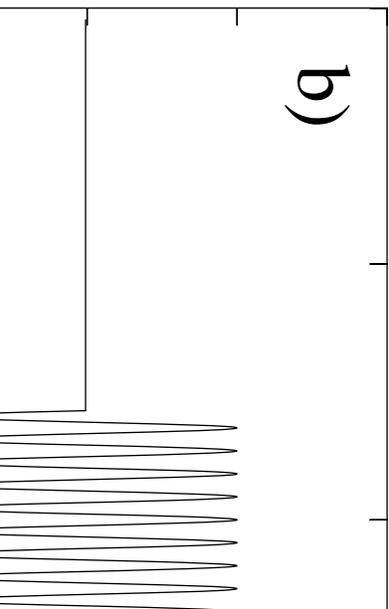
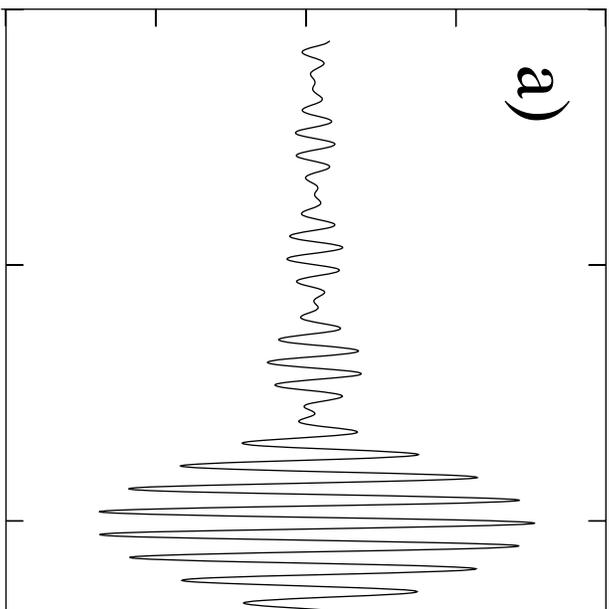
\end{document}